\documentclass[a4paper,11pt]{article}
\pdfoutput=1 

\usepackage{jheppub} 
\usepackage{hyperref}
\usepackage[T1]{fontenc} 
\usepackage{amsmath,amssymb}
\usepackage{slashed}
\usepackage{tikz}
\usepackage{upgreek}
\usetikzlibrary{arrows,positioning,cd,calc,math,decorations.pathreplacing,decorations.markings,decorations.pathmorphing
}

\tikzset{wave/.style={decorate, decoration=snake}}

\newcommand{\tr}{\text{tr\,}}

\newcommand{\Tau}{\mathcal{T}}
\newcommand{\res}{\text{Res\,}}
\newcommand{\diag}{\text{diag}}

\newcommand{\pd}{\partial}
\newcommand{\bs}{\boldsymbol}
\newcommand{\mc}{\mathcal}
\newcommand{\eq}[1]{\begin{equation}\begin{gathered}#1\end{gathered}\end{equation}}

\newcommand{\Fhyp}{{}_2F_1}

\title{$\mathcal{N}=2^*$ gauge theory,  free fermions on the torus and Painlev\'e VI}

\author[a,b,c]{Giulio Bonelli,}
\author[a,b,c]{Fabrizio Del Monte,}
\author[d,e,f]{Pavlo Gavrylenko}
\author[a,b,c]{and Alessandro Tanzini}

\affiliation[a]{International School of Advanced Studies (SISSA),
via Bonomea 265, 34136 Trieste, Italy}
\affiliation[b]{Institute for Geometry and Physics (IGAP),
via Beirut 2/1, 34151 Trieste, Italy}
\affiliation[c]{INFN, Sezione di Trieste}
\affiliation[d]{Center for Advanced Studies, Skolkovo Institute of Science and Technology, Nobel Street 1, 121205 Moscow, Russia}
\affiliation[e]{NRU HSE, International Laboratory of Representation Theory and Mathematical Physics, Usacheva 6, 119048 Moscow, Russia}
\affiliation[f]{Bogolyubov Institute for Theoretical Physics, Metrologichna 14-b,  03143 Kyiv, Ukraine}

\emailAdd{bonelli@sissa.it}
\emailAdd{fdelmont@sissa.it}
\emailAdd{pasha.145@gmail.com}
\emailAdd{tanzini@sissa.it}

\abstract{In this paper we study the extension of Painlev\'e/gauge theory correspondence to circular quivers by focusing on the special case of $SU(2)$ $\mathcal{N}=2^*$ theory. 
We show that the Nekrasov-Okounkov partition function of this gauge theory provides an explicit combinatorial expression and a Fredholm determinant formula for the tau-function describing isomonodromic deformations of $SL_2$ flat connections on the one-punctured torus. This is achieved by reformulating the Riemann-Hilbert problem associated to the latter in terms of chiral conformal blocks of a free-fermionic algebra.
This viewpoint provides 
the exact solution of the renormalization group flow of the 
$SU(2)$ $\mathcal{N}=2^*$ theory on self-dual $\Omega$-background and, in the Seiberg-Witten limit, an elegant relation between the IR and UV gauge couplings.
}

\begin{document} 

\maketitle
\flushbottom

\section{Introduction}

Four-dimensional supersymmetric gauge theories with eight supercharges can be studied with a variety of complementary approaches,
deeply rooted in the algebraic properties of their BPS states. A central r\^ole in this game is played by classical \cite{Seiberg:1994rs} and quantum \cite{Nekrasov:2009rc} integrable systems
and two-dimensional conformal field theories \cite{Alday:2009aq,Nekrasov:2015wsu}. 
In this context it has been observed that renormalization group
flows of supersymmetric theories in the self-dual $\Omega$-background can be studied as isomonodromic deformations of Hitchin connections \cite{Bonelli:2016qwg}.
In the special case of $SU(2)$ gauge theories, this problem reduces to the study of Painlev\'e equations. 
More precisely, the isomonodromic problems related to Painlev\'e equations are identified with the Hitchin system in the oper limit, 
the associated tau functions are computed for the full set of isomonodromic problems associated to Painlev\'e confluence diagram and compared
with topological string partition functions on the relevant local Seiberg-Witten curve geometries. 
These computations involve not only topological strings at large radius but also in other phases e.g. conifold point \cite{Bonelli:2016qwg}.
This line of research was triggered by previous work devoted to the solution of some Painlev\' e equations in terms of instanton combinatorics and two-dimensional conformal blocks of Virasoro algebra, beginning with \cite{Gamayun:2012ma}. Further developments along these lines are reported in 
\cite{Gamayun:2013auu,Iorgov:2014vla,Bershtein:2014yia,Gavrylenko:2016moe,Gavrylenko:2018ckn,Nagoya:2015cja,nagoya2018remarks}. This correspondence has been then broadened to the case of q-difference Painlev\'e equations, q-Virasoro algebra \cite{Bershtein:2016aef,
Bershtein:2017swf,Bershtein:2018srt,Mironov:2017sqp,jimbo2017cft,matsuhira2018combinatorial} and five dimensional $\mathcal{N}=1$ gauge theories and nonperturbative topological strings \cite{Grassi:2014zfa,Bonelli:2016idi,Bonelli:2017ptp,Bonelli:2017gdk,Grassi:2018spf}.

The four-dimensional theories in question can be seen as arising as the world-volume theories of stacks of M5 branes wrapped around a punctured Riemann surface $C_{g,n}$ described by the compactification of the relevant 6d $\mathcal{N}=(0,2) $ superconformal field theory \cite{Gaiotto:2009we}. This latter theory is fully determined by specifying a Lie algebra $\mathfrak{g}$, so that a four-dimensional class $S$ theory is determined by the set of data $(\mathfrak{g},C_{g,n} ,D)$. The additional data $D$ specifies half-BPS codimension 2 defects at the punctures of $C$, or equivalently, in the zero-area limit of the compactification, boundary conditions at the punctures \cite{Gaiotto:2009hg}. In this context, AGT correspondence can be seen as a suitable quantization of Hitchin's integrable system \cite{Bonelli:2009zp,Teschner:2010je,Bonelli:2011na,Vartanov:2013ima}.
If regular punctures only are considered the gauge theory is superconformal and the data $D$ are given by the residues at the punctures of the meromorphic differential specifying the corresponding Hitchin system. Their gauge theory counterpart are the masses and representations of the matter hypermultiplets. 
The flow to asymptotically free theories and/or strongly coupled fixed points of Argyres-Douglas type is realized by including 
also irregular punctures (higher order poles) \cite{Gaiotto:2009ma, Marshakov:2009gn, Bonelli:2011aa}. In these cases further data have to be specified at the punctures in the form of Stokes' parameters. 
As we will recall later, see \eqref{eq:RHPSphere} and \eqref{eq:LinSysTorus}, the isomonodromy problem can be formulated starting from a Riemann-Hilbert problem specified by the asymptotic behaviour of the Hitchin connection at the punctures. 
%
%
%
The isomonodromic problems studied so far in relation to Painlev\'e concern only linear quiver gauge theories, which correspond to CFT on the two-sphere. The goal of this work is to show how the correspondence between Painlev\'e equations, CFT and gauge theories can be extended to circular quivers: this amounts to defining the CFT on the torus and including hypermultiplets in the adjoint representation in the gauge theory. We wish to illustrate this by considering explicitly the simplest case of the one-punctured torus and $SL_2$, which corresponds to the mass-deformation of $\mathcal{N}=4$ SYM known as $\mathcal{N}=2^*$.

The main result of this paper is to give an explicit realisation of the tau-function for the isomonodromic deformations of a flat $SL_2$-connection
with a single simple pole on the torus in terms of Nekrasov-Okounkov dual partition functions $Z^D$ for the $SU(2)$ $\mathcal{N}=2^*$ theory \cite{Nekrasov:2003rj}:
\begin{equation}
\Tau_{gauge}(\tau)=\frac{Z^D(\eta,a,m,\tau)}{Z_{twist}(\tau)^2}.
\end{equation}	
We also get two alternative expressions for the tau function in terms of a Fourier series of Virasoro conformal blocks with only integer or half-integer shifts:
\eq{\label{eq:TauFuncVirasoro}
\eta(\tau)^{-1}\theta_2(2Q|2\tau)\mc T_{gauge}(\tau)=\sum_{n\in\mathbb Z}e^{i(n+\frac12)\eta} \tr_{\mathcal{V}_{a+n+\frac 12}}(q^{L_0}V_m(0))=Z^D_{1/2}(\eta,a,m,\tau)\\
\eta(\tau)^{-1}\theta_3(2Q|2\tau)\mc T_{gauge}(\tau)=\sum_{n\in\mathbb Z}e^{in\eta} \tr_{\mathcal{V}_{a+n}}(q^{L_0}V_m(0))=Z^D_{0}(\eta,a,m,\tau).
}
Here $Q$ is the solution of the isomonodromic system on torus which we specify in Sect. {\ref 3}, 
$\mc V_{a+n/2}$ is the Virasoro Verma module with dimension $(a+n/2)^2$, $V_m(0)$ is a Virasoro primary field inserted on the torus at the origin in cylindrical coordinates, and $q=e^{2\pi i\tau}$.
The general statement for higher rank and arbitrary number of regular punctures will be discussed in an upcoming paper \cite{Bonelli:2018ToApp}.

Finally, let us comment on the connection of these results with Seiberg-Witten theory and integrable systems: the Hitchin's connection we use to
define the system \eqref{eq:LinSysTorus} is a deautonomization of the Lax matrix of the elliptic Calogero-Moser system. More generally, we are considering the dependence on the marginal deformations of the class $S$ theory by deforming the Lax matrix that describes its Coulomb branch, similarly to what has been done in Seiberg-Witten theory by means of Whitham deformations 
\cite{Gorsky:1995zq,Levin:1997tb,Edelstein:1999xk,Bonelli:2009zp,Teschner:2010je}. 
The Hamiltonians of the integrable system become time-dependent, and the times are such marginal deformations.

In the following Section {\ref 2} we briefly recall the known results on the two-sphere.
In Sect. {\ref 3} we illustrate the new results obtained
for the one-punctured torus case, namely we discuss the free fermion realization
of the Riemann-Hilbert problem (RHP) kernel and the corresponding isomonodromic tau function.
In Sect. {\ref 4} we provide a Fredholm determinant formula for the latter and discuss gauge theory/topological string implications of our results.
We report our conclusions in Sect. {\ref 5}.
The remaining third of the paper consists of technical appendices.
 In particular, in Appendix {\ref{sec:Elliptic} ,\ref{sec:Fusion}} we set up notations that are used throughout the paper; in Appendix \ref{sec:Periodicity}, \ref{sec:SelfConst} we provide some consistency checks, while in Appendix \ref{eq:AppNum} we study the asymptotic behavior of our solution in the $\tau\rightarrow+i\infty$ limit and compute the corresponding monodromies in the spirit of \cite{19821137}. Finally, in Appendix \ref{App:DetForm} we provide some complementary determinantal formulae.

\section{Review of Painlev\'e/CFT correspondence on the sphere}
\label{2}

In this section we give a brief overview of the recent results on Conformal Field Theory approach to isomonodromic deformations.
The problem we are interested in concerns the variations at fixed masses of the Hitchin connection $L$ associated to the four-dimensional gauge theory 
obtained from M5-branes compactification on $C_{g,n}$. In this section we review the genus zero case.
The starting point is a system of linear ODEs
\begin{equation}\label{eq:LinSysSphere}
\begin{cases}
\partial_z Y(z)=L(z)Y(z), 	\\
Y(z_0)=\mathbb{I}_N.
\end{cases}
\end{equation}
where both $Y$ and $L$ are $N\times N$ matrices.  
In general, $L(z)$ can be allowed to have meromorphic entries, but we will restrict ourselves to the case of simple poles at points $z_1,\dots,z_n$\footnote{In the following we will only deal with so-called regular singularities, i.e. simple poles. This means that the solution to the Riemann-Hilbert problem will involve monodromies but not Stokes' jumps, and the corresponding gauge theory is conformal. Stokes' phenomena occur when $L(z)$ has higher order poles, and the corresponding gauge theory is asymptotically free or at an Argyres-Douglas point.}
\begin{equation}\label{eq:LaxSphere}
L(z)=\sum_{k=1}^n\frac{A_k}{z-z_k},
\end{equation}
where $A_k$ are constant matrices. The problem of finding the solution $Y(z)$ of this linear system can be rephrased as the following Riemann-Hilbert Problem (RHP): find a matrix valued, holomorphic and multivalued function on $\mathbb{P}^1\setminus\{z_k\}$ with the following properties:
\begin{equation}\label{eq:RHPSphere}
\begin{cases}
Y(z\sim z_k)=G_k(z)(z-z_k)^{\boldsymbol\theta_k}C_k, \\
\det Y(z)\ne 0, & z\ne z_k, \\
Y(z_0)=\mathbb{I}_N,
\end{cases}
\end{equation}
where $\boldsymbol\theta_k=\diag(\theta_1,\dots,\theta_N) $ is the diagonal matrix of eigenvalues\footnote{Subject to the non-resonance condition \cite{Jimbo:1981zz}.} of $A_k$, that is diagonalized at $z_k$ by holomorphic matrix-valued functions $G_k(z)$:
\begin{equation}
\boldsymbol\theta_k=G_k^{-1}(a_k)A_kG_k(a_k).
\end{equation}
The monodromies acquired by $Y(z)$ upon analytic continuation around $z_k$ are given by 
\begin{equation}
M_k=C_k^{-1} e^{2\pi i\boldsymbol\theta_k}C_k
\end{equation}
because of this, the $\theta_k$'s are called local monodromy exponents. The problem of finding a solution to the linear system \eqref{eq:LinSysSphere} is completely equivalent to that of finding a solution to  the Riemann-Hilbert Problem \eqref{eq:RHPSphere}, a particular instance of the so-called Riemann-Hilbert correspondence (see e.g. \cite{fokas2006painleve,conte2011painleve}).

Further, it is possible to study deformations of such linear system, and in particular to study the ones that keep the monodromies of the solution fixed. Such deformations are called isomonodromic, and they are given by a set of flows with times $t_k$, whose Hamiltonians $H_k$ are generated by the so-called isomonodromic tau function $\Tau$, first defined in \cite{Jimbo:1981zz,Jimbo:1982zz} and then generalized in \cite{Malgrange1982,2010CMaPh.294..539B}
\begin{equation}\label{eq:TauFunctSphere1}
\partial_{t_k}\log\Tau=H_k=\frac{1}{4\pi i}\oint_{\gamma_k}\tr L^2(z)dz=\frac{1}{2}\text{Res}_{z=z_k}\tr L^2(z),
\end{equation}
where $\gamma_k$ is a small loop around $z_k$. In general, for Fuchsian $SL_2$ isomonodromic problems on higher genus Riemann surfaces, the number of isomonodromic times is equal to the dimension of the moduli space of genus $g$ curves with $n$ punctures
$\dim\mathcal{M}_{g,n}=3g-3+n$, and 
the formula above gets generalized in terms of overlap of the quadratic differential $\tr L^2$ with the relevant Beltrami differentials.

In genus zero the first nontrivial case is the four punctured sphere, for which the isomonodromic deformation equations take the form of Painlev\'e VI. On the torus, the first nontrivial case is that with one puncture, and the isomonodromic time is the modular parameter $\tau$. The corresponding Hamiltonian is an integral over the A-cycle
\begin{equation}\label{eq:TauHamilt}
H_\tau=\frac{1}{2} \oint_A \tr L^2(z) dz=2\pi i\partial_\tau\log\Tau .
\end{equation}

It was already noted in the late '70s \cite{1978223,1979201,1979577,1979871,1980531} that two-dimensional Quantum Field Theory provides a useful framework to solve such a class of problems. In fact, the result of those papers have been recently extended and simplified by using the much more powerful tools developed in the past few decades in the context of two-dimensional Conformal Field Theory \cite{Gamayun:2012ma,Iorgov:2014vla,Gavrylenko:2016moe,Gavrylenko:2015wla}.

In order to find a solution to the RHP \eqref{eq:RHPSphere}, one has to find a function $Y(z)$ with prescribed monodromies, singular behavior and normalization. The CFT that engineers the solution in the case of an $N\times N$ system is a chiral CFT with $W_N$ symmetry \cite{Gavrylenko:2018ckn}, but to make the essential points clearer in this paper we will consider the $2\times 2$ case, that is solved by the more familiar conformal theories with Virasoro symmetry \cite{Iorgov:2014vla}. Because of this, from now on unless otherwise stated it will be implicitly assumed everywhere that the matrices are $2\times 2$ and the CFT is the chiral half of Liouville theory.

A natural candidate for a CFT description of the function $Y$ is a holomorphic conformal block: more precisely, one has to consider Liouville theory at $c=1$, and a chiral conformal block with insertion at the location of the poles of primary fields $V_k$ with Liouville charge $\theta_k$,
weight
\begin{equation}
\Delta_k=\theta_k^2
\end{equation}
and normalization~\footnote{comparing to \cite{Iorgov:2014vla} we change $\theta_k\to-\theta_k$}
\begin{equation}\label{eq:NormalizVert}
\begin{gathered}
\langle \sigma|V_k|\sigma'\rangle\equiv N(\sigma,\theta_k,\sigma') \\
=\frac{G(1+\sigma-\sigma'+\theta_k)G(1+\sigma'-\sigma+\theta_k)G(1-\theta_k-\sigma-\sigma')G(1+\sigma+\sigma'-\theta_k)}{G(1+2\theta_k)G(1+2\sigma)G(1-2\sigma')}
\end{gathered}
\end{equation}
where $G$ is the Barnes' G-function. The fields $\phi_s,\tilde{\phi}_s$, with $s=\pm$, represent the two fusion channels
\begin{equation}\label{eq:DegFusion}
[\phi_{(1,2)}]\times[V_a]=[V_{a-1/2}]+[V_{a+1/2}]
\end{equation}
of the degenerate field $\phi_{(1,2)}$ of Liouville charge $1/2$, inserted at $z,z_0$. The chiral block one has to consider is then
\begin{equation}
\Phi_{rs}\left(z,z_0;\{\theta_k\},\{\sigma_k\}\right)=\frac{\langle V_1(z_1)\dots V_n(z_n)\tilde{\phi}_r(z_0)\phi_s(z)\rangle}{\langle V_1(z_1) \dots V_n(z_n) \rangle}
\end{equation}
where $\sigma_k$ are the internal weights of the conformal block. To avoid cluttering of indices we adopt the matrix notation
\begin{equation}\label{ob1}
\Phi(z,z_0;\{\theta_k\},\{\sigma_k\})=\frac{\langle V_1(z_1)\dots V_n(z_n)\tilde{\phi}(z_0)\otimes\phi(z)\rangle}{\langle V_1(z_1) \dots V_n(z_n) \rangle}.
\end{equation}

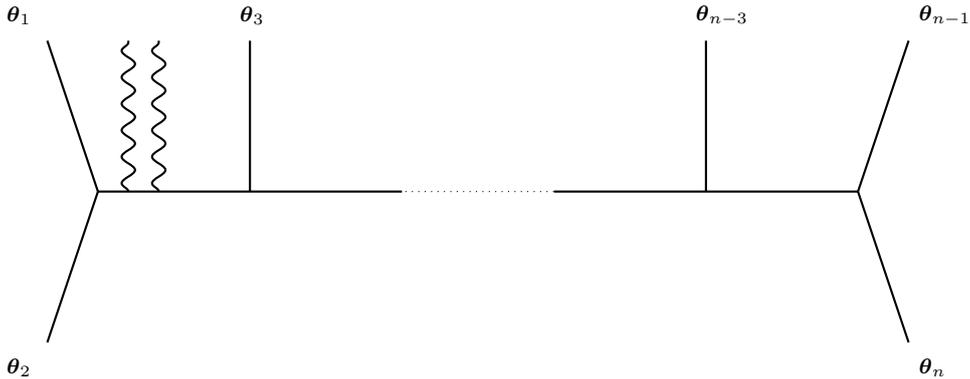
\begin{figure}[h!]
\begin{center}
\begin{tikzpicture}

\tikzmath{\scale=2;\radius=1;\offset=1/3;}
\begin{scope}[scale=\scale]

\draw[thick](0,\radius)--(\offset,0);
\draw[thick](0,-\radius)--(\offset,0);
\draw[thick](\radius+\offset,0)--(\offset,0);
\draw[thick,wave](\offset+0.2,0)--(\offset+0.2,\radius);
\draw[thick,wave](\offset+0.4,0)--(\offset+0.4,\radius);
\draw[thick](\radius+\offset,0)--(\radius+\offset,\radius);
\draw[thick](\radius+\offset,0)--(2*\radius+\offset,0);
\draw[dotted](3*\radius+\offset,0)--(2*\radius+\offset,0);
\draw[thick](3*\radius+\offset,0)--(4*\radius+\offset,0);
\draw[thick](4*\radius+\offset,\radius)--(4*\radius+\offset,0);
\draw[thick](5*\radius+\offset,0)--(4*\radius+\offset,0);
\draw[thick](5*\radius+\offset,0)--(5*\radius+2*\offset,\radius);
\draw[thick](5*\radius+\offset,0)--(5*\radius+2*\offset,-\radius);
\node[anchor=west] at(-\offset,\radius+\offset/2) {\scriptsize $\bs \theta_1$};
\node[anchor=west] at(-\offset,-\radius-\offset/2) {\scriptsize $\bs \theta_2$};
\node[anchor=west] at(-\offset/2.5+\offset+\radius,\radius+\offset/2) {\scriptsize $\bs \theta_3$};
\node[anchor=west] at(-\offset/2.5+\offset+4*\radius,\radius+\offset/2) {\scriptsize $\bs \theta_{n-3}$};
\node[anchor=west] at(2*\offset+5*\radius,\radius+\offset/2) {\scriptsize $\bs \theta_{n-1}$};
\node[anchor=west] at(2*\offset+5*\radius,-\radius-\offset/2) {\scriptsize $\bs \theta_n$};
\end{scope}
\end{tikzpicture}
\end{center}

\caption{$n$ point sphere conformal block with degenerate field insertions}
\label{fig:SphereBlock}
\end{figure}

The singular behavior of $\Phi$  is determined by the OPE of the fields, and its monodromies can be computed through the Moore-Seiberg formalism \cite{moore1989} (see \cite{Alday:2009aq,Drukker:2009id,Okuda:2014fja} for a recent review in application to supersymmetric gauge theories), which we quickly recap in Appendix \ref{sec:Fusion}. Since the fusion kernel for a degenerate field and a primary is known explicitly, the analytic continuation can be fully computed.

The chiral block \eqref{ob1}, however, does not yet have nice monodromy transformations. Indeed, because of the fusion rules of degenerate fields with primaries \eqref{eq:DegFusion}, the internal weights of the conformal block get shifted when performing a fusion operation. For the construction of the solution to the RHP from Virasoro conformal block in \cite{Iorgov:2014vla} it was crucial to show that even though a single fusion operation shifts the weights by half-integers, to perform a complete cycle on the sphere one has to perform each fusion twice, so that monodromy operations in the CFT at genus zero yield in general integer shifts of internal weights. We will see in the next section that this is not true in higher genus and we will show how to overcome this problem on the torus.

In order to diagonalize the shift operator, one has to consider the Fourier series
of conformal blocks
\begin{equation}
\Phi^D(z,z_0;\{\theta_k\},\{\sigma_k\})=\sum_{\textbf{n}\in\mathbb{Z}^{n-3}}e^{i\textbf{n}\cdot\boldsymbol\eta}\Phi(z,z_0;\{\theta_k\},\{\sigma_k+n_k\}),
\end{equation}
where D stands for "dual". The monodromies of  $\Phi^D$ are computable by means of the Moore-Seiberg formalism and are constructed from the CFT's fusion and braiding matrices: the expressions are rather involved, especially in the case of many punctures, and we refer to the original papers for their explicit form \cite{Iorgov:2014vla,Gavrylenko:2015wla}. $\Phi^D$  has the prescribed monodromies and singular behavior around $z_k$ and to identify it with the normalized solution to the RHP, it is still necessary to remove the additional singularity at $z=z_0$ coming from the OPE of the two degenerate fields. The normalized solution is then
\begin{equation}
Y(z;z_0)\equiv Y^{-1}(z_0)Y(z)=(z-z_0)^{1/2}\Phi^D(z,z_0;\{\theta_k\},\{\sigma_k \}),
\end{equation} 
Finally, one can obtain the tau function from $Y$ by expanding the above expression for $z\sim z_0$: by using
\begin{align}
L(z)=Y^{-1}(z)\partial_z Y(z), && \tr L(z)=0
\end{align}
we can write
\begin{equation}
\frac1{z-z_0}\tr Y^{-1}(z_0)Y(z)=\frac{2}{z-z_0}+\frac{z-z_0}{2}\tr L^2(z)+\dots.
\end{equation}
On the CFT side, the expansion can be done by using the OPE of $\phi,\tilde{\phi}$ in \eqref{ob1}, and equating order by order in $z-z_0$ we get
\begin{equation}\label{eq:SphereHamilt}
\begin{split}
\frac{1}{2}\tr L^2(z) & =\frac{\langle V_1(z_1)\dots V_n(z_n)T(z) \rangle_D}{\langle V_1(z_1)\dots V_n(z_n) \rangle}\\
& =\sum_{k=1}^n\left[\frac{\Delta_k}{(z-z_k)^2}+\frac{1}{z-z_k}\partial_{z_k}\log\langle V_1(z_1)\dots V_n(z_n) \rangle \right].
\end{split}
\end{equation}
By using \eqref{eq:TauFunctSphere1} and taking the residues in $z_k$ of the above expression, we get that the tau function is simply the Fourier transformed chiral conformal block of primary fields
\begin{equation}\label{eq:TauFunctSphere}
\Tau=\langle V_1(z_1)\dots V_n(z_n) \rangle_D.
\end{equation}

It is also possible to construct a solution of the linear system \eqref{eq:LinSysSphere} by using free fermions \cite{Iorgov:2014vla,Gavrylenko:2016moe}. In this case, the normalization factor is $z-z_0$, as it must compensate the pole of the fermion propagator. 
In this way, we see that we can regard the Fourier transformed conformal block as a kernel in the variables $z,z_0$,
\begin{equation}\label{eq:KernelSphere}
\Psi^D(z,z_0;\{\theta_k\},\{\sigma_k\})=\frac{Y^{-1}(z_0)Y(z)}{z-z_0}
\end{equation}
where now $\Psi$ is a free fermionic conformal block, and $\Psi^D$ its Fourier series as in the degenerate field case. In fact, the description using free fermions seems the most natural, since the Fourier transform in this case simply comes from the decomposition of the free fermionic Hilbert space. The tau function has the same expression as in the previous case.

Note that because the tau function can be written as a correlator of primary fields in a free fermionic CFT, it coincides, up to a factor coming from the normalization of the vertex operators, with Nekrasov-Okounkov dual partition function \cite{Nekrasov:2003rj} for a linear quiver gauge-theory, with quiver diagram given by the conformal block as in Figure \ref{fig:SphereBlock}, but without the wiggly lines. If one considers instead the purely Liouville representation using degenerate fields, in order to reach the same conclusion one is forced to use the AGT correspondence \cite{Alday:2009aq}. This identification has been studied in greater detail in the case of four punctures corresponding to the sixth Painlev\'e equation and to the $N_f=4$ theory, in \cite{Gamayun:2013auu,Bonelli:2016qwg}, where also the degeneration of PVI to other Painlev\'e equations has been studied and identified with holomorphic decoupling of hypermultiplets or flow to strongly coupled Argyres-Douglas points in the gauge theory.

\section{Isomonodromic deformations and torus conformal blocks}
\label{3}

\subsection{Linear systems on the torus}
To generalize the Lax matrix \eqref{eq:LaxSphere} to the case of the torus, we must take into account the Riemann-Roch theorem. Because of it, there is no function with only a single simple pole on the torus, and in general a Lax matrix $L(z)$ required to have simple poles at given points will transform nontrivially along the $A$ and $B$ cycles\footnote{In general, $L$ transforms as a connection, so in the transition functions $T_A,T_B$ there could be also a nonhomogeneous term. However, these matrices can be chosen so that they are $z$-independent up to a scalar multiple \cite{Levin:2013kca}.}:
\begin{align}\label{eq:ConnTwist}
L(z+1)=T_AL(z)T_A^{-1} , && L(z+\tau)=T_B L(z)T_B^{-1} ,
\end{align}
where the twists $T_A,T_B$ satisfy
\begin{equation}
T_AT_B^{-1}T_A^{-1}T_B=\zeta,
\end{equation}
where
\begin{equation}
\zeta=e^{2\pi i c_1/N},
\end{equation}
and $c_1=0,\dots, N-1$ is the first Chern class of the bundle with the centre of $SL(N)$ as structure group, which classifies the inequivalent flat bundles on the torus \cite{Levin:2013kca}. One can go from one bundle to the other by means of singular gauge transformations, known as Hecke transformations \cite{Levin:2001nm}, so that it is without loss of generality that we will deal with the case $c_1=0$, corresponding to the Lax matrix of the N-particles elliptic Calogero-Moser system. This is known to describe isomonodromic deformations on the one-punctured torus, with isomonodromic time $\tau$ \cite{levin1999hierarchies,takasaki1999elliptic}. It is also the Lax matrix of the integrable system describing the Seiberg-Witten theory of four-dimensional $SU(N)$ super Yang-Mills with one hypermultiplet in the adjoint representation of the gauge group, or $\mathcal{N}=2^*$ theory \cite{DHoker:1997hut,DHoker:1999yni,DHoker:2002kfd}.

The $SL(2,\mathbb{C})$ linear system \eqref{eq:LinSysSphere} with one simple pole at $z=0$ on the torus is then
\begin{align}\label{eq:LinSysTorus}
\begin{cases}
\partial_z Y(z|\tau) =L(z|\tau) Y(z|\tau), \\
Y(z_0|\tau)=\mathbb{I}_2,
\end{cases}
 &&
L(z|\tau) = \left( \begin{array}{cc}
p & m x(2Q,z) \\
mx(-2Q,z) & -p
\end{array} \right),
\end{align}
\begin{align}
T_A=\mathbb{I}_2 && T_B= e^{2\pi i \bs{Q}}, && \zeta=1
\end{align}
where
\begin{equation}
x(u,z)=\frac{\theta_1(z-u|\tau)\theta_1'(\tau)}{\theta_1(z|\tau)\theta_1(u|\tau)},
\end{equation}
and we also used the notation~\footnote{
Here and below we use the standard Pauli matrices
$\sigma^x=\begin{pmatrix}0&1\\1&0\end{pmatrix}$,
$\sigma^y=\begin{pmatrix}0&-i\\i&0\end{pmatrix}$,
$\sigma^z=\begin{pmatrix}1&0\\0&-1\end{pmatrix}$.
}
\begin{equation}
e^{2\pi i\bs{Q}}=e^{2\pi iQ\sigma^z}.
\end{equation}
As a consequence of \eqref{eq:ConnTwist}, the solution $Y$ will have, besides the usual monodromies acting on the right, also twists acting on the left:
\begin{align}\label{eq:YTransfTorus}
Y(\gamma_A\cdot z|\tau)=Y(z)M_A, && Y(\gamma_B\cdot z|\tau)=e^{2\pi i \bs{Q}}Y(z)M_B, && Y(\gamma_k\cdot z|\tau)=Y(z)M_k.
\end{align}
Differently from the monodromies, the twists are not constant along the isomonodromic flows. In this case with one puncture, we can easily use identity \eqref{eq:LameWP} and the defining equation to compute the isomonodromic Hamiltonian
\begin{equation}
\begin{split}
H_\tau & =\frac{1}{2}\oint_A\tr L^2(z)dz=\int_0^1 dz\left[p^2-m^2\left(\wp(2Q|\tau)-\wp(z|\tau)\right) \right] \\
& = p^2-m^2\wp(2Q|\tau)-2m^2\eta_1(\tau),
\end{split}
\end{equation}
associated to the time $2\pi i\tau$. The last term comes from
\begin{equation}
\int_0^1dz\wp(z|\tau)dz=-\int_0^1dz\zeta'(z|\tau)=\zeta(0)-\zeta(1)=-2\eta_1(\tau),
\end{equation}
and since it is a function of $\tau$ only it does not contribute to the isomonodromy deformation equations, which are the Hamilton equations for this Hamiltonian\footnote{This equation is the isomonodromy deformation equation because it is equivalent to the Lax pair equation
\begin{equation}\label{eq:ZeroCurv}
2\pi i\partial_\tau L+\partial_z M+[M,L]=0,
\end{equation}
which can be shown, by using equation \eqref{eq:LameProp0}, to be the zero-curvature compatibility condition of the system
\begin{equation}\label{eq:LaxPairSystem}
\begin{cases}
\partial_zY=LY, \\
2\pi i\partial_\tau Y=-MY,
\end{cases}
\end{equation}
where $M$ is the other matrix of the Lax pair of $L$:
\begin{equation}
M= m\left(\begin{array}{cc}
\wp(2Q) & \partial_Q x(2Q,z) \\
\partial_Q x(-2Q,z) & \wp(2Q)
\end{array}\right).
\end{equation}
} and take the form of a special case of Painlev\'e VI for $Q$ \cite{levin1999hierarchies,takasaki1999elliptic}:
\begin{equation}
(2\pi i)^2\frac{d^2Q}{d\tau^2}=m^2\wp'(2Q).
\label{eq:deaut_Calogero}
\end{equation}
We see that the twist is the Painlev\'e transcendent, and as such is a function of $\tau$ that in general cannot be expressed in terms of usual special functions.

\subsection{Monodromies of torus conformal blocks}

\begin{figure}[h!]
\begin{center}
\begin{tikzpicture}

\tikzmath{\scale=0.95;\radius=1;\length=1.4;\hlength=1;\ind=0.2;}


\begin{scope}[scale=\scale]
\begin{scope}
\draw[rounded corners=1cm*\scale, thick] (0,\radius)--(-\radius-\hlength,\radius)--(-\radius-\hlength,-\radius)--
(\radius+\hlength,-\radius)--(\radius+\hlength,\radius)--(0,\radius);
\draw[dashed,->] (0,\length+\radius+0.1) ..controls (0.4,\length+\radius+0.3) and (0.4,\length+\radius+0.3).. (0.9,\length+\radius+0.1);
\draw[thick](0.75,\radius)--+(0,\length);
\draw[thick,wave](0,\radius)--+(0,\length);
\draw[thick,wave](-0.75,\radius)--+(0,\length);
\draw[->](2.5,0)--(3,0);
\node[anchor=west] at(-0.2,-\radius+0.25) {\scriptsize $\bs a$};
\node[rotate=-45,anchor=west] at(0.2,\radius) {\scriptsize $\bs a - \frac{s-s'}{2}$};
\node[rotate=-45,anchor=west] at(-0.55,\radius) {\scriptsize $\bs a -\frac{s}{2}$};
\end{scope}

\begin{scope}[xshift=5.5cm]
\draw[rounded corners=1cm*\scale, thick] (0,\radius)--(-\radius-\hlength,\radius)--(-\radius-\hlength,-\radius)--
(\radius+\hlength,-\radius)--(\radius+\hlength,\radius)--(0,\radius);
\draw[rounded corners=1.2cm*\scale,dashed,<-] (-0.9,\radius+\ind)--(-\radius-\hlength-\ind,\radius+\ind)--(-\radius-\hlength-\ind,-\radius-\ind)--
(\radius+\hlength+\ind,-\radius-\ind)--(\radius+\hlength+\ind,\radius+\ind)--(1.1,\radius+\ind);
\draw[thick](0,\radius)--+(0,\length);
\draw[thick,wave](0.75,\radius)--+(0,\length);
\draw[thick,wave](-0.75,\radius)--+(0,\length);
\draw[->](2.5,0)--(3,0);
\node[,anchor=west] at(-0.2,-\radius+0.25) {\scriptsize $\bs a $};
\node[rotate=-45,anchor=west] at(0.2,\radius) {\scriptsize $\bs a - \frac{s''}{2}$};
\node[rotate=-45,anchor=west] at(-0.55,\radius) {\scriptsize $\bs a -\frac{s}{2}$};
\end{scope}


\begin{scope}[xshift=11cm]
\draw[rounded corners=1cm*\scale, thick] (0,\radius)--(-\radius-\hlength,\radius)--(-\radius-\hlength,-\radius)--
(\radius+\hlength,-\radius)--(\radius+\hlength,\radius)--(0,\radius);
\draw[thick,wave](0,\radius)--+(0,\length);
\draw[dashed,->] (-0.75,\length+\radius+0.1) ..controls (-0.35,\length+\radius+0.3) and (-0.35,\length+\radius+0.3).. (0.1,\length+\radius+0.1);
\draw[thick](0.75,\radius)--+(0,\length);
\draw[thick,wave](-0.75,\radius)--+(0,\length);
\node[anchor=west] at(-0.3,-\radius+0.25) {\scriptsize $\bs a-\frac{s''}{2} $};
\node[rotate=-45,anchor=west] at(0.2,\radius) {\scriptsize $\bs a - \frac{s}{2}$};
\node[rotate=-45,anchor=west] at(-0.55,\radius) {\scriptsize $\bs a$};
\end{scope}

\end{scope}
\end{tikzpicture}
\end{center}

\caption{B-cycle monodromy for the one-punctured torus}
\label{Fig:BCycle}
\end{figure}
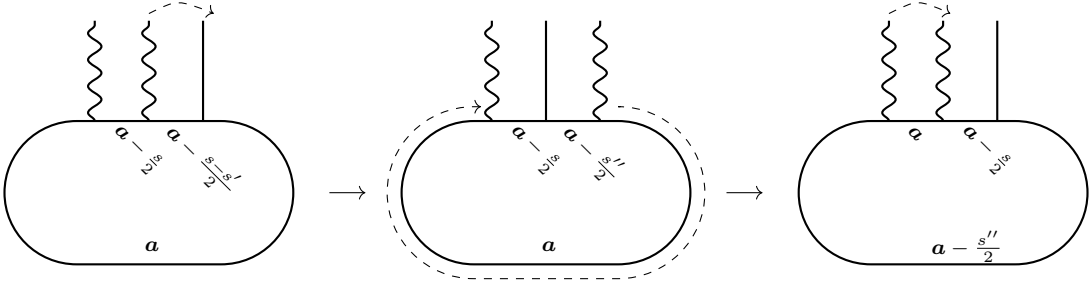

We will now turn to the main result of the paper: the derivation from CFT of the solution to the linear system \eqref{eq:LinSysTorus} and the tau function of the isomonodromic problem. To this end, we consider the chiral block
\begin{equation}
\Phi(z,z_0|\tau,a,m)=\langle V_{m}(0)\tilde{\phi}(z_0)\otimes\phi(z) \rangle = \frac{1}{Z(\tau)}\tr_{\mathcal{V}_a} \left( q^{L_0}V_m(0)\tilde{\phi}(z_0)\otimes\phi(z)\right)
\end{equation}
in Liouville Conformal Field Theory at $c=1$, where as in the previous section $\phi_i,\tilde{\phi}_i$ are degenerate fields, $V_m$ is a primary field with Liouville charge $m$, and weight $\Delta_m=m^2$, and $Z(\tau)$ is the partition function of the CFT. Because of the results already available on the sphere, the object
\begin{equation}
\Phi^D(z,z_0)=\sum_n e^{in\eta}\Phi(z,z_0;\tau,m,a+n)
\end{equation}
has prescribed monodromies in conjugacy classes~\footnote{The factor of $i$ in $M_A$ comes from the Jacobian of transition from plane to cylinder for a field of dimension $\frac14$.}
\begin{align}
M_A\sim i e^{2\pi i\bs a}, && M_1\sim e^{2\pi i\bs m}.
\end{align}
As was stressed in the previous section, in the case of the torus we do not always have only integer shifts of the internal weights when transporting $z$ around a closed loop: when we move the degenerate field around the B-cycle of the torus, we perform fusion with the primary $V_m$ only once, so that because of the fusion rules \eqref{eq:DegFusion} the internal weights get shifted by half-integers, as is shown in Figure \ref{Fig:BCycle}. As a consequence, the Fourier transform $\Phi^D$, that includes only integer shifts, will not transform into itself under B-cycle monodromy. Let us make this observation more precise by computing how $\Phi$ transforms when the degenerate field goes in a loop around the B-cycle.

The transformation is given in Figure \ref{Fig:BCycle}, which means that
\begin{equation}
\Phi_{ss'}(\gamma_B\cdot z)=i\sum_{s''=\pm}\Phi_{ss''}(z)e^{-\frac{s''}{2}\overleftarrow{\partial_a}}e^{i\pi s''(a-s/2)}F_{s''s'}(a-s/2,m,a)e^{-i\pi s'a}
\end{equation}
where we denoted by $e^{-\frac{s''}{2}\overleftarrow{\partial_a}}$ the operator that acts on the left by shifting $a\rightarrow a-s''/2$ (operators must act from the right because such is the action of the monodromies).
Now we use the property
\begin{align}
F_{s''s'}(\bs a-s/2,m,a)=ss''F_{s''s'} (a,m-1/2,a),
\end{align}
of the fusion matrix, which can be easily checked given the explicit expression of the fusion matrix in Appendix \ref{sec:Fusion}, together with 
\begin{equation}
e^{i\pi s''(a-s/2)}=-ss''ie^{i\pi s''a}
\end{equation}
to rewrite the above monodromy action as a matrix action (operator valued, because of the shifts):
\begin{equation}\label{eq:BCycleDeg}
\Phi(\gamma_B\cdot z)=\Phi(z)e^{-\frac{1}{2}\overleftarrow{\partial_a}\sigma^z}e^{i\pi\bs a}F(a,m-1/2,a)e^{-i\pi\bs a}.
\end{equation}

At first glance, it would seem that due to the half-integer shifts in the intermediate channel, in the case of the torus one should consider the Fourier transform
\begin{equation}
\sum_n e^{\frac{in\eta}{2}}\Phi(z,z_0;\tau,m,a+n/2),
\end{equation}
as it diagonalizes the shift operator $e^{-\frac{1}{2}\overleftarrow{\partial}_{\bs a}}$. However, doing so the monodromy action along the  A-cycle is spoiled, since
\begin{gather}
\Phi(\gamma_A\cdot z,a+n)=i\Phi(z,a+n)e^{-2\pi i\bs a} , \\ \Phi(\gamma_A\cdot z, a+n+1/2)=-i\Phi(z,a+n+1/2)e^{-2\pi i\bs a} .
\end{gather}

This can be remedied by considering the CFT with an additional $U(1)$ boson as in \cite{Iorgov:2014vla}, so that we have free fermions instead of degenerate fields:
\begin{align}
\label{eq:fermions_def}
\psi(z)\equiv e^{i\varphi(z)}\phi(z), && \bar{\psi}(z)\equiv e^{-i\varphi(z)}\tilde{\phi}(z).
\end{align}
The presence of the $U(1)$ boson shifts all monodromy exponents by the eigenvalue $\sigma$ of the zero-mode of $\partial\varphi$: we might as well set $\sigma=0$, however as we will see, it will turn out to be meaningful to keep this factor for a while. The effect of adding the $U(1)$ boson is that the additional sign factor along the A-cycle cancels out with that of the degenerate fields, so that the free fermion conformal block
\begin{equation}\label{eq:FermionCorr}
\begin{split}
\Psi^D(z,z_0;\tau,m,a,\sigma,\eta,\rho) & \equiv\sum_{n,k} e^{\frac{in\eta}{2}}e^{4\pi i(\rho+1/2)(n/2+k)}\Psi(z,z_0;m,a+n/2,\sigma+1/2+n/2+k)\\
&\equiv \langle V_m(0)\bar{\psi}(z_0)\otimes\psi(z) \rangle
\end{split}
\end{equation}
has numerical monodromies along all noncontractible cycles. In this definition the shifts of the Virasoro highest weight $a\rightarrow a+n/2$ and of the Heisenberg charge ($\sigma+1/2\rightarrow\sigma+1/2+n/2+k$) correspond to the shift of the fermionic charge(s) of two-component fermions by $(n+k, k)$.
By using the above considerations, together with expression \eqref{eq:BCycleDeg} for the B-cycle monodromy, and the action
\begin{equation}
\Psi^D(z,z_0;\tau,m,a,\sigma)e^{-\frac{1}{2}\overleftarrow{\partial_a}\sigma^z}e^{-\frac{1}{2}\overleftarrow{\partial_\sigma}}
=\Psi^D(z,z_0;\tau,m,a,\sigma)e^{i\frac{\eta}{2}\sigma^z+2\pi i\rho},
\end{equation}
 we see that the monodromies around the A- and B-cycles of the torus take the form
\eq{\label{eq:TorusMonodromies}
\hat M_A=e^{-2\pi i\bs a-2\pi i\sigma}\equiv e^{-2\pi i\sigma} M_A,\\
\hat M_B=e^{i\frac{\eta}{2}\sigma^z+2\pi i\rho} e^{i\pi \bs a}F(a,m-1/2,a)e^{-i\pi \bs a}\equiv e^{2\pi i\rho}M_B.
}

Above, $M_A,M_B$ are the part of the conformal block monodromies $\hat{M}_A,\hat{M_B}$ that are independent from the additional $U(1)$ charges $\sigma,\rho$, and they have unit determinant. As we will see in the next section, they are the monodromy matrices of the linear system solution $Y$. \footnote{
Recall that $\det Y(z)=1$, so that the extra $U(1)$ factors $e^{2\pi i\rho}$ and $e^{-2\pi i\sigma}$ from the point of view of the linear system are introduced artificially. In fact, they are arbitrary and we can set them to any value, but it turns out to be convenient to keep them arbitrary throughout the computations.
}

Let us make some final remarks on this computation. The A-cycle monodromy is encoded in the mode expansion of the complex fermions:
\eq{
\psi(z)=\sum_{p\in \mathbb Z+\frac12}\psi_p e^{2\pi i(p-\bs a-(\sigma+1/2))z}
}
which comes from the change of coordinate from the plane to the torus $w=e^{-2\pi iz}$. We see that the $1/2$ shift of $\sigma$ must be added in order to cancel anti-periodicity of the natural mode expansion. In the same way, in the computation of the B-cycle monodromy we shifted by $1/2$ the parameter $\rho$ in order to cancel the $(-1)$ factor coming from fermions re-ordering \footnote{These two shifts mean that we are fermionizing the degenerate fields into fermions which are periodic along both cycles on the torus (in the sense that no additional signs are involved in the computation of monodromies). The shift in $\sigma$ amounts to the periodicity condition on the cylinder, while that in $\rho$ is implemented in the operator formalism we are using by an insertion of $(-)^F$ in all our traces.}.

The signs of the shifts by $n$ are defined by the expression for $L_0$, twisted by two fermionic charges ${\bf H}_1$ and ${\bf H}_2$:
\eq{
L_0=const + L_0^{(0)}+{\bf H}_1 (\sigma+a) + {\bf H}_2(\sigma-a),
}
where 
\begin{align}
[L_0^{(0)},\psi_{i,p}]=-n\psi_{i,p}, && [{\bf H}_i,\psi_{j,p}]=\delta_{ij}\psi_{j,p}.
\end{align}

\subsection{CFT solution to the Riemann-Hilbert problem and tau function}

To relate the free fermion correlator \eqref{eq:FermionCorr} to the solution $Y$ of the linear system \eqref{eq:LinSysTorus}, we need to take  care of two things: the fact that the correlator has an additional simple pole in $z=z_0$ with residue one that $Y$ does not have, and the fact that $Y$ has, in addition to the monodromies acting from the right, twists acting from the left, as in \eqref{eq:YTransfTorus}. The generalization of the sphere kernel \eqref{eq:KernelSphere} to a kernel on the torus is then:
\begin{equation}\label{eq:KernelTorus}
Y^{-1}(z_0)\Xi(z-z_0)Y(z)=\frac{\langle V_m(0)\bar{\psi}(z_0)\otimes\psi(z)\rangle}{\langle V_m(0) \rangle},
\end{equation} 
where we defined the matrix
\begin{equation}
\Xi(z)\equiv \diag\left(x(\sigma\tau+\rho-Q,z),x(\sigma\tau+\rho+Q,z) \right) e^{-2i\pi\sigma z}
\end{equation}
whose transformations along the two cycles of the torus are such that it cancels the twists of the solution $Y$, while also giving the additional shifts due to the $U(1)$ boson charge $\sigma$.
The two sides of the equations have the same singularities with same singular behavior, and same monodromies both in $z$ and $z_0$. Because of this, they coincide.
It is also useful to introduce new notations
\eq{
\widetilde Q_1=-\sigma\tau-\rho+Q,\quad \widetilde Q_2=-\sigma\tau-\rho-Q,\quad \widetilde{\bs Q}=\diag(\widetilde Q_1,\widetilde Q_2).
}


As in the case of the sphere, we now expand both sides of the equation to obtain the tau function. In particular, we need only to expand the trace of \eqref{eq:KernelTorus}
\begin{equation}
\sum_\alpha\frac{\langle\psi_\alpha(z)\bar\psi_\alpha(z_0) V_m(0)\rangle}{\langle V_m(0)\rangle}=-\tr\left[Y(z)Y^{-1}(z_0)\Xi(z-z_0)\right].
\end{equation}
The expansion of the first two factors is:
\begin{equation}
Y(z)Y^{-1}(z_0)=\left(\mathbb{I}+(z-z_0)L(z_c)+\frac{(z-z_0)^2}{2}L^2(z_c) \right),
\end{equation}
where $z_c=\frac{z+z_0}2$. Then, we expand the theta functions in the diagonal matrix:
\begin{equation}
\frac{\theta_1'(\tau)}{\theta_1(z-z_0|\tau)}=\frac{1}{z-z_0}-\frac{z-z_0}{6}\frac{\theta_1'''(\tau)}{\theta_1'(\tau)}+O((z-z_0)^3) ,
\end{equation}
\eq{
e^{-2\pi i\sigma(z-z_0)}\frac{\theta_1(z-z_0+\widetilde{\bs Q}|\tau)}{\theta_1(\widetilde{\bs{Q}}|\tau)}=1+(z-z_0)\left(\frac{\theta_1'(\widetilde{\bs Q}|\tau)}{\theta_1(\widetilde{\bs Q}|\tau)}-2\pi i\sigma\right)+\\+\frac{(z-z_0)^2}{2}\left(\frac{\theta_1''(\widetilde{\bs Q}|\tau)}{\theta_1(\widetilde{\bs Q}|\tau)}-\frac{\theta_1'(\widetilde{\bs Q}|\tau)}{\theta_1(\widetilde{\bs Q}|\tau)}4\pi i\sigma+(2\pi i\sigma)^2\right) +O((z-z_0)^3)
}

Putting everything together, we find that the $O(z-z_0)$ term in the expansion above is
\eq{
\label{eq:T_def}
\sum_\alpha\frac{\langle\frac12 :\partial\bar\psi_\alpha(z_c)\psi_\alpha(z_c)+\partial\psi_\alpha(z_c)\bar\psi_\alpha(z_c) : V_m(0)\rangle}{\langle V_m(0)\rangle}=
\frac{\langle T(z_c) V_m(0)\rangle}{\langle V_m(0)\rangle}=
\\=
\frac{1}{2}\tr\left(L^2(z_c)+\frac{\theta_1''(\widetilde{\bs Q}|\tau)}{\theta_1(\widetilde{\bs Q}|\tau)}+2(L(z_c)-2\pi i\sigma)\frac{\theta_1'(\widetilde{\bs Q}|\tau)}{\theta(\widetilde{\bs Q}|\tau)}+(2\pi i\sigma)^2-\frac{1}{3}\frac{\theta_1'''(\tau)}{\theta_1'(\tau)} \right).
}

There are some additional terms with respect to what we found for the sphere. However they can be rearranged in a more convenient form. We use the fact that the diagonal part of the Lax matrix \eqref{eq:LinSysTorus} consists of the momenta of the Hamiltonian system \cite{Levin:2013kca} to write
\begin{equation}
\begin{split}
&\tr\left(\frac{\theta_1''(\widetilde{\bs Q}|\tau)}{\theta_1(\widetilde{\bs Q}|\tau)}+2(L(z_c)-2\pi i\sigma)\frac{\theta_1'(\widetilde{\bs Q}|\tau)}{\theta(\widetilde{\bs Q}|\tau)}+(2\pi i\sigma)^2-\frac{1}{3}\frac{\theta_1'''(\tau)}{\theta_1'(\tau)} \right)=\\
& =\sum_{i=1}^2\left(\frac{\theta_1''(\widetilde Q_i)}{\theta_1(\widetilde Q_i|\tau)}+2(p_i-2\pi i\sigma)\frac{\theta_1'(\widetilde Q_i|\tau)}{\theta_1(\widetilde Q_i|\tau)}+(2\pi i\sigma)^2-\frac{1}{3}\frac{\theta_1'''(\tau)}{\theta_1(\tau)} \right).
\end{split}
\end{equation}
We now use the heat equation for $\theta_1$, as well as the relation for the coordinates and momenta of the nonautonomous system
\begin{align}
\theta_1''=4\pi i\partial_\tau\theta_1, && p_i=2\pi i\partial_\tau Q_i,
\end{align}
to write the above expression as
\begin{equation}
\begin{split}
& 4\pi i\sum_{i=1}^2\left[\frac{\theta'_{1,\tau}(\widetilde Q_i|\tau)}{\theta_1(\widetilde Q_i|\tau)}+\partial_\tau(Q_i-\sigma\tau-\rho)\frac{\theta_1'(\widetilde Q_i|\tau)}{\theta_1(\widetilde Q_i|\tau)}+(2\pi i\sigma)^2-\frac{1}{3}\frac{\partial_\tau\theta'(\tau)}{\theta_1(\tau)} \right]= \\
& =4\pi i\sum_{i=1}^2\partial_\tau\log e^{\pi i\tau\sigma^2}\frac{\theta_1(\widetilde Q_i|\tau)}{\theta_1'(\tau)^{1/3}}=4\pi i\partial_\tau\log \left(e^{2\pi i\tau\sigma^2}\frac{\theta_1(\widetilde Q_1|\tau)}{\eta(\tau)}\frac{\theta_1(\widetilde Q_2|\tau)}{\eta(\tau)} \right)= \\
& =4\pi i\partial_\tau\log\left(Z_{twist}(Q,\rho,\sigma,\tau)\right),
\end{split}
\end{equation}
where in the last equality we noted that the argument of the logarithm is the partition function of two free complex twisted fermions with the twists defined by the Lax connection \eqref{eq:LinSysTorus} with extra $U(1)$ shift. Plugging the latter formula into \eqref{eq:T_def} we find
\begin{equation}\label{eq:trL2}
\frac{1}{2}\tr L^2(z_c)+2\pi i\partial_\tau\log\left(Z_{twist}(Q,\rho,\sigma,\tau) \right)=\frac{\langle V_m(0) T(z_c)\rangle}{\langle V_m\rangle}.
\end{equation}
We can use the Virasoro Ward identity for an energy-momentum tensor insertion on the torus, which takes the form
\begin{equation}
\begin{split}\label{eq:VirasoroWardTorus}
\frac{\langle T(z) V_m(0) \rangle}{\langle  V_m\rangle} & =\langle T \rangle+m^2\left[\wp(z|\tau)+2\eta_1(\tau) \right]+2\pi i\partial_\tau\log\langle  V_m\rangle.
\end{split}
\end{equation}
In order to identify the tau function from this equation, we will use a general result from \cite{Levin:2013kca}, in which the authors study isomonodromic deformation problems on elliptic curves. The relation between $\tr L^2$ and the isomonodromic Hamiltonian, specified to the case of one puncture, is
\begin{equation}\label{eq:TorusHamilt}
\frac{1}{2}\tr L^2(z)=H_\tau+m^2 (\wp(z)+2\eta_1).
\end{equation}
\label{eq:Hamiltonian1}
$m^2$ is the Casimir of the orbit around the puncture, and the isomonodromic Hamiltonian turns out to be
\begin{equation}
H_\tau=2\pi i\partial_\tau\log\Tau=2\pi i\partial_\tau\log\frac{\langle V_m\rangle}{Z_{twist}(\tau)},
\end{equation}
so that the correlation function is
\eq{\label{eq:TauFunctTorus}
Z^D(\tau)=\langle V_m\rangle = \sum_{n,k\in\mathbb Z}e^{4\pi i(\rho+1/2)(n/2+k)}e^{\frac{in\eta}{2}} \tr_{\mathcal{V}_{a+n/2}\otimes \mc{F}_{\sigma+1/2+n/2+k}}(q^{L_0}V_m(0)) =\\= Z_{twist}(\tau) \Tau(\tau)=
e^{2\pi i\tau\sigma^2}\eta(\tau)^{-2} \theta_1(\sigma\tau+\rho+Q(\tau))\theta_1(\sigma\tau+\rho-Q(\tau)) \mc T(\tau).
}

The tau function is then the correlator of primary fields in a free fermionic chiral  CFT, as in the case of the sphere, but instead of being normalized by the partition function of the CFT itself, it is normalized with the partition function of twisted complex fermions, with twists given by the isomonodromic problem under consideration. This result is more general, and holds for the case of the $n$-punctured torus, and for more general semi-degenerate $N\times N$ linear systems, solved by $W_N$ Conformal Field Theories \cite{Bonelli:2018ToApp}.

In fact, note that this is nothing else but the free fermion expression of the dual gauge theory partition function $Z^D$ that appears in the original paper by Nekrasov and Okounkov \cite{Nekrasov:2003rj}, so that
{\begin{equation}\label{eq:TauFunctDualPart}
\Tau(\eta,a,m,\tau)=\frac{G(1+m)^2}{G(1+2m)} \frac{Z^D(\eta,a,m,\rho,\sigma,\tau)}{Z_{twist}(\eta,a,m,\rho,\sigma,\tau)}.
\end{equation}
Where the additional $m$-dependent factor comes due to our normalization of the vertices, as we show below. In fact, since we are defining the isomonodromic tau function by the property
\begin{equation}
2\pi i\partial_\tau\log\Tau=H_\tau,
\end{equation}
constant multiplicative factors are irrelevant, and we may as well consider the tau function to be
\begin{equation}
\Tau_{gauge}(\tau)\equiv\frac{Z^D(\tau)}{Z_{twist}(\tau)}.
\end{equation} }

At this moment we exploit the arbitrary $U(1)$ charge to establish relations between the Fourier series of Virasoro conformal blocks, which we will call Virasoro dual partition functions, and the full dual partition function, computed over the free fermionic Hilbert space. First we expand this latter over its pure Virasoro and Heisenberg contributions:
\eq{
Z^D(\tau)=\eta(\tau)^{-1}\sum_{n,k\in\mathbb Z}e^{i\eta n/2} e^{4\pi i(\rho+1/2)(k+n/2)}q^{(\sigma+1/2+k+n/2)^2}\tr_{\mathcal{V}_{a+n/2}}\left(q^{L_0} V_m(0)\right)=\\=
Z^D_0(\tau)\eta(\tau)^{-1} e^{2\pi i\tau\sigma^2}\sum_{k\in\mathbb Z} e^{2\pi i\tau (k+1/2)^2} e^{4\pi i(k+1/2)(\sigma\tau+\rho+1/2)}+\\+
Z^D_{1/2}(\tau)\eta(\tau)^{-1} e^{2\pi i\tau\sigma^2}\sum_{k\in\mathbb Z} e^{2\pi i\tau k^2} e^{4\pi ik(\sigma\tau+\rho+1/2)}=\\=
-Z^D_0(\tau) e^{2\pi i\tau\sigma^2}\eta(\tau)^{-1}\theta_2(2\sigma\tau+2\rho|2\tau)+Z^D_{1/2}(\tau) e^{2\pi i\tau\sigma^2}\eta(\tau)^{-1}\theta_3(2\sigma\tau+2\rho|2\tau).
\label{eq:tau0_tau12}
}
The above equation defines $Z_0^D$ and $Z_{1/2}^D$, which are the Fourier transform of the one-point torus conformal block containing respectively only integer or half-integer shifts:
\eq{
Z_{\epsilon/2}^D(\eta,a,m,\tau)=\sum_{n\in\mathbb Z+\frac\epsilon2}
e^{in\eta} \tr_{\mathcal{V}_{a+n}}(q^{L_0}V_m(0)).
}
 The trace over the Fock space has been resummed and yields the theta function and Dedekind eta factors. Notice further that the variables $\sigma$ and $\rho$ enter the above formula only through the two theta functions. Now we use the addition formula for theta functions
\eq{
\theta_1(x-y|\tau)\theta_1(x+y|\tau)=\theta_3(2x|2\tau)\theta_2(2y|2\tau)-\theta_2(2x|2\tau)\theta_3(2y|2\tau)
}
and rewrite the relation between dual partition function and isomonodromic tau function \eqref{eq:TauFunctTorus} in a form
\eq{
Z^D(\tau) =
\mc T(\tau)e^{2\pi i\sigma^2\tau}\eta(\tau)^{-2}\bigl(
-\theta_2(2\sigma\tau+2\rho|2\tau)\theta_3(2Q|2\tau)+\theta_3(2\sigma\tau+2\rho|2\tau)\theta_2(2Q|2\tau)\bigr).
}
Comparing the two formulas we find two relations, free from $\sigma$ and $\rho$:
\eq{
Z^D_0(\tau)=\eta(\tau)^{-1}\theta_3(2Q|2\tau)\mc T(\tau)\,,\\
Z^D_{1/2}(\tau)=\eta(\tau)^{-1}\theta_2(2Q|2\tau)\mc T(\tau)\,.
\label{eq:ZD-tau}
}
One consequence of these formulas is the relation
\eq{
\label{eq:solution_isomonodromic}
\frac{\theta_3(2Q|2\tau)}{\theta_2(2Q|2\tau)}=\frac{Z_0^D(\tau)}{Z_{1/2}^D(\tau)}
}
which allows us to express solution $Q$ of the isomonodromic system in terms of CFT/gauge theory objects. Another	possibility is to use a "minimal choice" for the extra charge of the $U(1)$ boson, setting $\sigma=\rho=0$. Then the above expression becomes
\begin{equation}
\Tau(\eta,a,m,\tau)=\frac{\eta(\tau)^2}{\theta_1(Q|\tau)^2}Z^D(\eta,a,m,0,0,\tau) .
\end{equation}

Let us see explicitly which objects of the free fermionic CFT yield the classical, perturbative and instanton part of the dual partition function. Expanding the trace in the basis of descendants $|a,\textbf{Y}\rangle $ one recovers the instanton expansion: the factor $q^{L_0}$ yields the classical partition function and the instanton counting parameter,
the normalization of the vertex operator \eqref{eq:NormalizVert} gives the perturbative contribution
\begin{equation}\label{nama}
N(a,m,a)=\frac{G(1+m)^2G(1-m-2a)G(1-m+2a)}{G(1+2m)G(1-2a)G(1+2a)}=\frac{G(1+m)^2}{G(1+2m)} Z_{pert}(a,m)
\end{equation}
together with the extra $m$-dependent factor in \eqref{eq:TauFunctDualPart}. Finally, the expansion of the conformal block itself in the basis of descendants labeled by partitions $|a,\textbf{Y} \rangle$ yields as usual the instanton contributions to the partition function~\footnote{
The precise statement is that to get Nekrasov factors one has to make $\res_0 L(z)dz$ of rank $1$ by appropriate $U(1)$ shift. It is the standard AGT trick: see also discussion in the end of Section~\ref{sec:determinant}. }.

\section{Gauge theory, topological strings and Fredholm determinants}\label{4}
\subsection{$\mathcal{N}=2^*$ gauge theory partition function as a Fredholm determinant }

\label{sec:determinant}

As we have shown in the previous subsection, in order to construct the solution of the RHP on the one-punctured torus in terms of CFT correlators, it is necessary to switch from degenerate fields to free fermions, see \eqref{eq:fermions_def}.
In \cite{Gavrylenko:2016moe} it was shown that in the four-punctured sphere case one can sum up the expression for the tau function into a single Fredholm determinant by means of the generalized Wick theorem for free fermions.
This determinant for the generic tau function on the sphere with $n$ punctures was also constructed in \cite{Gavrylenko:2016zlf} in a mathematically rigorous way. This was then shown to satisfy the  Jimbo-Miwa-Ueno definition of the tau function and to reproduce the expansion of the relevant Nekrasov partition function.
More recent understanding of such determinant formulas in the sphere case, together with a simplified proof, can be found in \cite{Cafasso:2017xgn}.

In \cite{Gavrylenko:2016zlf} the RHP is solved by decomposing the $n$-punctured sphere in trinions or pairs of pants, which are three-punctured spheres, thus reducing the RHP on the sphere with $n$ punctures to the problem of properly gluing solutions to RHPs on three-punctured spheres, with punctures at $0,1,\infty$. These are given, normalized by their asymptotics in zero, by
\eq{
Y_0(w)=(1-w)^{(m-\gamma)}\times\\\times
\begin{pmatrix}
{}_2F_1(m,m+2a,2a,w)&\frac{-mw}{2a-1}{}_2F_1(1+m,1+m-2a,2-2a,w)\\
\frac{m}{2a}{}_2F_1(1+m,m+2a,1+2a,w)&{}_2F_1(m,1+m-2a,1-2a,w)
\end{pmatrix},
}
where ${_2F_1}$ are hypergeometric functions, and $\gamma$ is a $U(1)$ shift about which we will comment later. Note that the solution above is essentially the same as \eqref{eq:AsymptoticSol} upon change of variables from spherical to cylindrical $w=e^{2\pi i z}$, where it appears in the study of the asymptotic behavior of the solution $Y$ on the torus.  The solution above is well-defined as a series in $z$, convergent for $|w|<1$, so we also define another normalized solution of the same problem, well-defined as a series in $w^{-1}$:
\eq{
Y_\infty(w)=\sigma^x Y_0(1/w) \sigma^x.
}

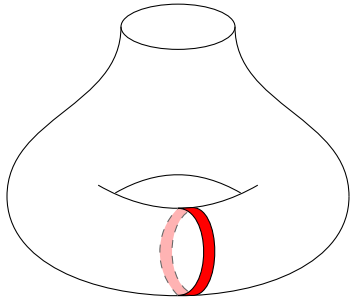
\begin{figure}[h!]
\begin{center}
\begin{tikzpicture}[scale=1.5]
\draw(0,0) circle[x radius=0.5, y radius =0.2];
\draw(-0.5,0)  to[out=270,in=90] (-1.5,-1.5) to[out=-90,in=-90] (1.5,-1.5);
\draw(0.5,0) to[out=270,in=90] (1.5,-1.5);

\draw(-0.7,-1.4) to[out= -30,in=210] (0.7,-1.4);
\draw(-0.55,-1.469) to[out= 30,in=-210] (0.55,-1.469);


\fill[red!30!white] (0,-2.375) to[out=135,in=225] (0,-1.605)
to (0.17,-1.6) to[out=190,in=170] (0.17,-2.373) --cycle;

\draw[dashed,color=black!60!white](0,-2.375) to[out=135,in=225] (0,-1.605)
to (0.17,-1.6) to[out=190,in=170] (0.17,-2.373) --cycle;

\fill[red](0,-2.375) to[out=0,in=0] (0,-1.605)
to (0.1,-1.6) to[out=10,in=-10] (0.1,-2.373) --cycle;

\draw(0,-2.375) to[out=0,in=0] (0,-1.605)
to (0.1,-1.6) to[out=10,in=-10] (0.1,-2.373) --cycle;

\end{tikzpicture}
\end{center}
\caption{Pants decomposition of one-punctured torus}
\label{fig:Torus}
\end{figure}

In this subsection, motivated by \cite{Gavrylenko:2016moe} and \cite{Gavrylenko:2016zlf}, we compute the relevant Fredholm determinant  by pants decomposition of the one-punctured torus as in Fig.~\ref{fig:Torus}. Namely, we expand the trace in \eqref{eq:TauFunctTorus} by inserting the identity operator on the free-fermion states, compute the matrix elements of $V_m$ by using the generalized Wick theorem, and finally arrive at the Fredholm determinant expression given below.

The one-punctured torus is obtained from the three-punctured sphere by gluing two legs of the trinion as in Figure \ref{fig:Torus}. We will take these to be the legs corresponding to be the punctures at $0,\infty$ in spherical coordinates, or $\pm i\infty$ in cylindrical coordinates. We thus have to show how this translates to an operation on the solution of the three-punctured RHP.

To glue the solutions defined at zero and infinity, we define the following integral kernels:
\eq{
\begin{aligned}
&{\sf a}(w,w')=D\frac{Y_0(qw)^{-1}Y_0(w')-\mathbb I}{q w-w'},&\quad
&{\sf b}(w,w')=-D\frac{Y_0(qw)^{-1}Y_\infty(w')}{qw-w'},&\\
&{\sf c}(w,w')=D^{-1}\frac{Y_\infty(w/q)^{-1}Y_0(w')}{w/q-w'},&\quad
&{\sf d}(w,w')=D^{-1}\frac{\mathbb{I}-Y_\infty(w/q)^{-1}Y_\infty(w')}{w/q-w'},&
\end{aligned}
}
where the diagonal matrix $D$ is given by the formula
\eq{
D=-q^{(1+\sigma)}e^{2\pi i\rho} \diag(q^{-a}e^{-2\pi i\beta},q^{a}e^{2\pi i\beta}),
}
with $e^{2\pi i\beta}$ given by
\eq{\label{eq:BetaSec3}
e^{2\pi i\beta}=e^{i\eta/2}\frac{\Gamma(1-2\alpha)\Gamma(2\alpha-m)}{\Gamma(2\alpha)\Gamma(1-2\alpha-m)}.
}

Let us now specify the Hilbert spaces on which the above operators act. Actually, there are at least two equivalent choices.
The first is to consider them as operators acting on the space of (row-)vector-valued functions
\eq{
\mc H=L^2(S^1)\otimes \mathbb C^2
}
on a circle $S^1=\{w, |w|=R\}$, where $|q|<R<1$. In this realisation, the action of an operator ${\sf A}$ on a function $f$ is defined by the integral
\eq{
({\sf A}f)(w)=\oint\frac{dw'}{2\pi i}f(w'){\sf A}(w',w).
}

We will instead employ a different description of this Hilbert space, which is better adapted for computational purposes and that makes contact naturally with the free fermion description: we will use the Fourier expansions of the kernels
\eq{
\label{eq:abcd_series}
\begin{aligned}
&{\sf a}(w,w')=\sum_{p,q\in \mathbb Z'_-}\frac{{\sf a}_{pq}}{w^{p+\frac12}w'^{q+\frac12}},&\quad&
&{\sf b}(w,w')=\sum_{p\in \mathbb Z'_-,q\in \mathbb Z'_+}\frac{{\sf b}_{pq}}{w^{p+\frac12}w'^{q+\frac12}}&\\
&{\sf c}(w,w')=\sum_{p\in \mathbb Z'_+,q\in \mathbb Z'_-}\frac{{\sf c}_{pq}}{w^{p+\frac12}w'^{q+\frac12}},&\quad&
&{\sf d}(w,w')=\sum_{p,q\in \mathbb Z'_+}\frac{{\sf d}_{pq}}{w^{p+\frac12}w'^{q+\frac12}}&,
\end{aligned}
}
where $\mathbb Z_+'=\{\frac12,\frac32,\frac53,\ldots\}$, $\mathbb Z_-'=\{-\frac12,-\frac32,-\frac52,\ldots\}$ are positive and negative half-integer spaces respectively, and $\mathbb Z'=\mathbb Z_-'\sqcup \mathbb Z_+'$.
In terms of Fourier modes we can describe our Hilbert space as $\mc H=\mathbb C^{\mathbb Z'}\otimes \mathbb C^2$, whose basis vectors are labelled by a pair $(p,\alpha)$ of one half-integer number $p\in\mathbb Z'$ and one matrix index $\alpha\in \{1,2\}$.
One can also define two subspaces of $\mc H$: $\mc H_+$, corresponding to $\mathbb Z_+'$ --- the subspace of non-negative Fourier modes, and $\mc H_-$, corresponding to $\mathbb Z_-'$ --- the subspace of negative Fourier modes.
We can easily see from \eqref{eq:abcd_series} that the operators ${\sf a, b, c, d}$ act non-trivially only between the following sub-spaces:
\eq{
{\sf a}: \mc H_+\to\mc H_-,\quad
{\sf b}: \mc H_-\to\mc H_-,\quad
{\sf c}: \mc H_+\to\mc H_+,\quad
{\sf d}: \mc H_-\to\mc H_+.\quad
}

Using the above definitions, together with the results of \cite{Gavrylenko:2016zlf}, one can write down the following expression for the  dual partition function $Z^D(\tau)$:
\eq{\label{eq:determinant}
Z^D(\tau)=N(a,m,a) q^{a^2+(\sigma+1/2)^2-1/12}e^{2\pi i(\rho+1/2)}\prod_{n=1}^\infty(1-q^n)^{-2\gamma^2}
\det\left(\mathbb{I}+K\right),
}
where the operator $K$ can be either written as a block matrix with respect to the decomposition $\mc H=\mc H_+\oplus \mc H_-$:
\eq{
K=
\begin{pmatrix}
{\sf c}&{\sf d}\\
{\sf a}&{\sf b}
\end{pmatrix}.
}
Alternatively, one can sum up the whole Fourier modes and define a single matrix integral kernel as
\eq{
K(w,w')={\sf a}(w,w')+{\sf b}(w,w')+{\sf c}(w,w')+{\sf d}(w,w').
}
acting on $L^2(S^1)\otimes \mathbb{C}^2$.



Let us now make some comments on \eqref{eq:determinant}. The factor $N(a,m,a)$, that we defined in \eqref{nama}, accounts both for the vertex normalization in CFT and the one-loop factor of the Nekrasov partition function, while $q^{a^2}$ is the classical contribution to the partition function. The Fredholm determinant ${\rm det} (1+K)$ is then identified with the instanton part of Nekrasov-Okounkov partition function for $\mathcal{N}=2^*$ gauge theory, up to a free fermion normalization depending on the background charges $(\gamma,\sigma,\rho)$, which are arbitrary and can be set to any value. Let us see how interesting results can be obtained by specializing these $U(1)$ charges to prescribed values.

First, note that the r.h.s. of \eqref{eq:determinant} does not depend on $\gamma$ since the $U(1)$ factor $\eta(q)^{-2\gamma^2}$ cancels the same contribution from the determinant. The advantage of having this extra shift $\gamma$ is that one can consider the two
cases $\gamma=0$ and $\gamma=m$~\footnote{
The first case corresponds to $\res_{w=1} L(w) dw\sim \diag(m,-m)$, whereas the second one corresponds to $\res_{w=1} L(w)dw\sim\diag (2m,0)$.
The second normalization was used in \cite{Nekrasov:2003rj}.}.
If $\gamma=0$, the summation over principal minors of $K$ gives Virasoro conformal blocks as in \eqref{eq:conf_block_AGT}, but in this case such minors have a complicated form.
If one puts instead $\gamma=m$, the minors of $K$ turn into factorized Nekrasov expressions: technically the $U(1)$ contribution in front of the determinant,
together with the $\eta(\tau)^{-1}$ from \eqref{eq:tau0_tau12}, cancels the $U(1)$ factor in \eqref{eq:conf_block_AGT}. For the explicit computations of minors see \cite{Gavrylenko:2016zlf}.

Let us consider now $\rho,\sigma$. Note that that dependence of $Z^D(\eta,a,m,\rho,\sigma,\tau)$ on these parameters, given by \eqref{eq:tau0_tau12}, is quite simple, as the only meaningful combination is $\sigma\tau+\rho$. Let us then put $\sigma=0$ and study the $\rho$-dependence. From \eqref{eq:TauFunctTorus} we can see that zeroes of $Z^{D}(\eta,a,m,\rho,\tau)$ in $\rho$ define the solution $Q(\tau)$ of the Painlev\'e VI equation \eqref{eq:deaut_Calogero}:
\eq{\label{eq:zero_of_tau}
Z^D(\eta,a,m,\pm Q(\tau)+k+l\tau,\tau)=0,
}
so that $Q$ may be found as zero of the Fredholm determinant \eqref{eq:determinant}. This formula is definitely the deautonomization of Krichever's formula \cite{Krichever1980} which gives coordinates of $N$ particles $Q_i(t)$ in the elliptic Calogero-Moser system as zeroes of the theta function: $\Theta(\vec U Q_i(t)+\vec V t+\vec W)=0$. Indeed, in Section \ref{sec:AutLimit} by studying explicitly the isospectral/autonomous limit, we find exactly Krichever's formula. The generalization of \eqref{eq:zero_of_tau} to $N>2$ will be given in \cite{Bonelli:2018ToApp}.

To explain the origin of \eqref{eq:zero_of_tau} in the spirit of \cite{Gavrylenko:2016zlf} we notice the following.
If one substitutes $\rho=Q(\tau)$, then the first row of $Y(w)$ has the following periodicity properties:
\eq{
Y_{1i}(w+1)=Y_{1i}(w),\quad Y_{1i}(w+\tau)=e^{2\pi i Q}Y_{1i}(w)M_B=Y_{1i}(w)\hat M_B.
}
We thus see that the two functions $Y_{11}(w)$ and $Y_{12}(w)$ are globally defined functions on the torus with prescribed monodromies $\hat{M}_A$ and $\hat{M}_B$, so it is possible to restrict them to the boundaries of the red strip in Fig.~\ref{fig:Torus} and therefore they belong to
the space of functions that have analytic continuations with prescribed mondromies inside and outside the red strip simultaneously.
This breaks the decomposition of the space of all functions into the space of functions analytic inside and outside the red strip, which holds in the generic position, and this is indicated by the vanishing of the determinant. We report some other identities that can be derived by setting the $U(1)$ charges to specific values in Appendix \ref{App:DetForm}. More details on the detailed proof of the results of this subsection will be reported elsewhere.

\subsection{Line operators and B-branes}

Let us briefly comment on the gauge/string theoretical interpretation of the Riemann-Hilbert kernel. We will first briefly review the interpretation of
the CFT construction in terms of Liouville degenerate vertex insertions on the sphere. As we showed before this construction has to be modified on higher genus Riemann surfaces by considering instead free fermions. Therefore, the construction that naturally arises in this case is rather a free fermion CFT, whose string/gauge theoretic interpretation will be outlined in the following. 

Theories of class $S$ have a canonical surface operator \cite{Gaiotto:2009fs}, given in the M-theory construction by an M2 brane embedded in spacetime $\mathbb{R}^4$, localized at a point $z\in C$. 
In the AGT correspondence, this is described by an insertion of a degenerate field of weight $\pm b/2$ at $z$. Wilson and 't Hooft loops living on the surface operator are computed by means of Verlinde loop operators using braiding and fusion of degenerate fields \cite{Alday:2009fs,Drukker:2009id}, in the same way as the monodromies of the fundamental solution of the linear system are computed in the CFT approach to isomonodromy equations.
When the Verlinde loop operator goes around an A-cycle of the Riemann surface, the monodromy operation acts multiplicatively on the conformal block. The corresponding line operator in gauge theory is a Wilson loop. When we go around a B-cycle, there is a shift of the internal Liouville momentum. In this case the line operator carries magnetic charge and thus it is a 't Hooft loop. 
A choice of A-cycle and B-cycles on the Riemann surface corresponds to a choice of S-duality frame in the gauge theory, and the modular group of the Riemann surface is the S-duality group. 

%
%

The Verlinde loop operators represent the braiding algebra of Wilson and 't Hooft loops at an operatorial level. 
The Fourier basis that is adopted in the isomonodromic setting is necessary in order to have an object that transforms linearly onto itself by monodromy so that we can identify the correlator with the fundamental solution of the linear system \eqref{eq:LinSysSphere} on the sphere. On this basis of the Hilbert space both 't Hooft loops and Wilson loops act by multiplication, which is possible because they are commuting operators. This basis can therefore be regarded as an S-duality complete basis of the Coulomb branch for loop operators, since Wilson and 't Hooft loops are treated on the same footing.

The situation is different in the case of the one-punctured torus, which was analyzed in  this paper. In our case, the Wilson and 't Hooft loop operators for $SU(2)$ anticommute, and by using only degenerate fields one is not able to construct a S-duality complete basis. Indeed,
on the torus the addition of an extra $U(1)$ boson in the CFT is required, leading to a free fermionic CFT. 
The dictionary between monodromy and gauge theory data is the following:
the monodromy at the puncture parametrises the mass of the adjoint hypermultiplet
\begin{equation}
\label{tr1}
{\rm Tr} M_1=2 {\rm cos}(2\pi m)\, ,
\end{equation}
the monodromy along the $A$-cycle parametrises the v.e.v. of the Wilson loop
in the fundamental representation
\begin{equation}
\label{trA}
{\rm Tr} M_A=2 {\rm cos}(2\pi a)
\end{equation}
in terms of the v.e.v. of the scalar field of the ${\mathcal N}=2$ vector multiplet
and 
the combined monodromy around the $A$ and the $B$ cycles 
\begin{equation}
\label{trAB}
{\rm Tr} M_AM_B=
\frac{1}{{\rm sin}(2\pi a)}\left[{\rm sin}\pi(2a-m)e^{-i(\eta/2-2\pi a)}
+
{\rm sin}\pi(2a+m)e^{i(\eta/2-2\pi a)}
\right]
\end{equation}
parametrises the v.e.v. of the minimal dyonic 't Hooft loop operator as computed in
\footnote{To compare with \cite{Drukker:2009id}, put $b=i$, $a\rightarrow ia$ remembering that our conformal weights are related to the Liouville charge by $\Delta_a=a^2$, while in \cite{Drukker:2009id} the usual convention $\delta_a=-a^2$ is used.}
\cite{Drukker:2009id}. From the view point of the Painlev\'e transcendent, $(a,\eta)$
are related to the initial conditions of $Q(\tau)$ as it is shown in Appendix \ref{eq:AppNum}, while from the Hitchin system perspective formulas \eqref{trA} and \eqref{trAB} together with
${\rm Tr} M_B$ are the Darboux coordinates of the moduli space of $SL(2,{\mathbb C})$ flat connections on the one punctured torus \cite{Nekrasov:2011bc}.

This CFT has a more natural interpretation in the topological string setting, where loop operators are computed in terms of brane amplitudes. Our proposal is that the isomonodromy deformation of the integrable system associated to the classical Seiberg-Witten curve is described in terms of topological B-brane amplitudes as discussed in \cite{Aganagic:2003qj}. There -- see sect. 4.7  and further elaborated in \cite{Dijkgraaf:2007sw,Dijkgraaf:2008fh} -- it is also suggested that for higher genus Riemann surfaces the most natural framework is given by considering the free-fermion grand-canonical partition function 
leading precisely to the Fourier basis and thus to the Nekrasov-Okounkov partition function. Indeed this latter can be regarded as a character of 
$\mathcal{W}_{1+\infty}$-algebra as in the topological B-brane setting of \cite{Aganagic:2003qj}. 

\subsection{The autonomous/Seiberg-Witten limit}\label{sec:AutLimit}

In this subsection we analyse the autonomous limit 
which gives the isospectral integrable system describing the Seiberg-Witten geometry of $\mathcal{N}=2^*$, namely Calogero-Moser. 
We proceed by finding the explicit solution of equation 
\eq{\label{eq:IsomonToIsosp}
H_\tau = (2\pi i\pd_\tau Q)^2-m^2(\wp(2Q|\tau)+2\eta_1(\tau))
}
in the scaling limit $H_\tau=\hbar^{-2}(u+O(\hbar))$, $m=\hbar^{-1}\mu$, for small variations of $\tau$, namely $\tau=\tau_0+\hbar t$, where $t\ll\hbar^{-1}$ in the limit $\hbar\to 0$~\footnote{
such small variations preserve the integrals of motion.
There is also another part of the problem: to find slow evolution of the integrals of motion at the time scale $t\sim\hbar^{-1}$.
The general approach to this problem, which gives rise to Whitham equations, is given in \cite{Krichever:2001cx}.
Relation of this approach to our general solution of the non-autonomous problem still has to be uncovered.}.
In this limit the scaled Hamiltonian is the Coulomb branch parameter
of the gauge theory. Indeed in the above limit we have
\begin{equation}\label{eq:SWCurve}
0=\det(L-\lambda\mathbb{I}_2)= H_\tau+m^2(\wp(2z)+2\eta_1(\tau))+\lambda^2\simeq\frac{1}{\hbar^2}\left[ u+\mu^2\left(\wp(2z)+2\eta_1(\tau)\right)+\tilde{\lambda}^2\right],
\end{equation}
which is the Seiberg-Witten curve for the $\mathcal{N}=2^*$ theory, so that the energy parameter is identified with the Coulomb branch modulus $u$\footnote{Note that all the quantities in the isomonodromic setting are dimensionless, being measured in Omega-background units.}. So in this limit \eqref{eq:IsomonToIsosp} takes the form of the energy conservation law
\eq{
\label{eq:energy}
u=(2\pi i\pd_t Q)^2-\mu^2 (\wp(2Q|\tau_0)+2\eta_1(\tau_0)).
}
As for any one-dimensional Hamiltonian system, we can integrate it by quadratures:
\eq{
t-t_0=\int^Q\frac{2\pi i dQ}{\sqrt{u+2\mu^2\eta_1(\tau)+\mu^2\wp(2Q|\tau_0)}},
}
however, to explicitly compute this integral we have to perform a couple of changes of variables. First we introduce the new variable
\begin{equation}
y=\frac{\theta_2(2Q|2\tau_0)}{\theta_3(2Q|2\tau_0)},
\end{equation}
that satisfies
\eq{
(2\pi i\pd_t y)^2=4 \pi^2 \theta_4(2\tau_0)^4 \frac{\theta_1(2Q|2\tau_0)^2\theta_4(2Q|2\tau_0)^2}{\theta_3(2Q|2\tau_0)^4}\left(u+2\mu^2\eta_1+\mu^2\wp(2Q|\tau_0)\right),
}
Where \eqref{eq:ThetaId1} has been used. Now substitute
\eq{
\wp(2Q|\tau_0)+2\eta_1(\tau_0)=-4\pi i\pd_{\tau_0}\log\theta_2(\tau_0)+
\left(\pi\theta_4(2\tau_0)^2\frac{\theta_2(2Q|2\tau_0)\theta_3(2Q|2\tau_0)}{\theta_1(2Q|2\tau_0)\theta_4(2Q|2\tau_0)}\right)^2.
}
Introducing
\begin{equation}
\tilde{u}=u-4\pi i \mu^2 \pd_{\tau_0}\log\theta_2(\tau_0)
\end{equation}
we rewrite \eqref{eq:energy} as
\eq{
(i\pd_t y)^2=\tilde{u} \theta_4(2\tau_0)^4\frac{\theta_1(2Q|2\tau_0)^2\theta_4(2Q|2\tau_0)^2}{\theta_3(2Q|2\tau_0)^4}+\pi^2\mu^2\theta_4(2\tau_0)^8\frac{\theta_2(2Q|\tau_0)^2}{\theta_3(2Q|\tau_0)^2}.
}
By using \eqref{eq:ThetaId2} we finally rewrite \eqref{eq:energy} as
\eq{
(i\pd_t y)^2=\tilde{u} \theta_2(2\tau_0)^2\theta_3(2\tau_0)^2\left(1-\frac{\theta_3(2\tau_0)^2}{\theta_2(2\tau_0)^2}y^2\right)\left(1-\frac{\theta_2(2\tau_0)^2}{\theta_3(2\tau_0)^2}y^2\right)+\\+
\pi^2\mu^2\theta_4(2\tau_0)^8 y^2.
}
We see that the problem is reduced to the computation of an elliptic integral. In order to do this, we introduce the new variable $\phi$ through
\eq{
y=\frac{\theta_2(2\phi|2\tau_{SW})}{\theta_3(2\phi|2\tau_{SW})},
}
where $\tau_{SW}$ is the complex modulus of the covering curve \eqref{eq:SWCurve}, that is the infrared gauge coupling of the $\mathcal{N}=2^*$ gauge theory. 
This is given by the polynomial in the r.h.s., and we get the expression
\eq{
(2\pi i\pd_t\phi)^2 \cdot \theta_2(2\tau_{SW})^2\theta_3(2\tau_{SW})^2\left(1-\frac{\theta_3(2\tau_{SW})^2}{\theta_2(2\tau_{SW})^2}y^2\right)
\left(1-\frac{\theta_2(2\tau_{SW})^2}{\theta_3(2\tau_{SW})^2} y^{2}\right)
=\\
=\tilde{u} \theta_2(2\tau_0)^2\theta_3(2\tau_0)^2\left(1-\frac{\theta_3(2\tau_0)^2}{\theta_2(2\tau_0)^2}y^2\right)\left(1-\frac{\theta_2(2\tau_0)^2}{\theta_3(2\tau_0)^2}y^2\right)+
\pi^2\mu^2\theta_4(2\tau_0)^8 y^2
}
To linearize the equation on $\phi$ we wish to cancel two bi-quadratic polynomials by solving this explicit equation on $2\tau_{SW}$:
\eq{\label{eq:SWUVCoupling}
\frac{\theta_2(2\tau_{SW})^2}{\theta_3(2\tau_{SW})^2}+\frac{\theta_3(2\tau_{SW})^2}{\theta_2(2\tau_{SW})^2}=
\frac{\theta_2(2\tau_0)^2}{\theta_3(2\tau_0)^2}+\frac{\theta_3(2\tau_0)^2}{\theta_2(2\tau_0)^2}-\frac{\mu^2}{\tilde{u}}\frac{\pi^2\theta_4(2\tau_0)^8}{\theta_2(2\tau_0)^2\theta_3(2\tau_0)^2}
}
The solution for $\phi$ is then given by the formula
\eq{
\phi=\frac{\sqrt{\tilde{u}}}{2\pi i}\frac{\theta_2(2\tau_0)\theta_3(2\tau_0)}{\theta_2(2\tau_{SW})\theta_3(2\tau_{SW})}t+\phi_0/2=\omega t+\phi_0/2.
}
Collecting together the two changes of variables we find that the coordinate $Q(t)$ should be found from the solution of the equation
\eq{\label{eq:AutonomousSol}
\frac{\theta_2(2Q(t)|2\tau_0)}{\theta_3(2Q(t)|2\tau_0)}=\frac{\theta_2(2\omega t+\phi_0)|2\tau_{SW})}{\theta_3(2\omega t+\phi_0)|2\tau_{SW})}.
}
This result has to be compared with \eqref{eq:solution_isomonodromic}: we see that in the isospectral limit dual partition functions can be effectively replaced by theta-functions. In fact, this formula coincides with the one in \cite{1999JMP....40.6339G}, expressing the exact solution of the elliptic Calogero-Moser model. As a byproduct, we found the explicit relation \eqref{eq:SWUVCoupling} between the UV coupling $\tau_0$ and the IR coupling $\tau_{SW}(\tau_0,\mu^2/u)$ for the $\mathcal{N}=2^*$ theory. 

In other terms, equation \eqref{eq:solution_isomonodromic} is the explicit solution of the renormalisation group flow of $\mathcal{N}=2^*$ theory in a self-dual $\Omega$-background, while
eq.\eqref{eq:SWUVCoupling} is the corresponding Seiberg-Witten limit.

The results above could be compared with the small mass expansion in terms of modular forms as found in \cite{Billo:2015pjb}. Actually, our finding suggests that modular anomaly equations are explicitly solved in terms of the corresponding isomonodromy problem, while their SW limit in terms of the corresponding Calogero-Moser system.

\section{Outlook and Conclusions}
\label{5}

In this paper we studied the extension of the Painlev\'e/gauge theory correspondence to circular quivers by focusing on the special case of $SU(2)$ $\mathcal{N}=2^*$ theory. 
We have shown that the Nekrasov-Okounkov partition function of this gauge theory provides an explicit combinatorial expression of the tau-function
of isomonodromic deformation problem for $SL_2$ flat connections on the one-punctured torus. We provided an explicit description of the initial conditions in terms on Darboux
coordinates of the moduli space of flat connections on the one punctured torus
and in terms of line operators in the $SU(2)$ $\mathcal{N}=2^*$ theory, see
equations \eqref{trA},\eqref{trAB}. 
The main tool we used is the construction of the fundamental solution of the associated Riemann-Hilbert problem in terms of conformal blocks of Virasoro plus $U(1)$
free-fermion algebra. 
This also allowed us to provide a Fredholm determinant formula for the 
tau functions/Nekrasov-Okounkov dual partition function.
With respect to the previously studied cases on the sphere,
a novelty arises due to the non-triviality of the flat bundle on the elliptic curve. This induces an extra factor which can be written as the partition function of twisted
free fermions on the torus, whose twisting parameters is given by the Painlev\'e transcendent of a special form of PVI equation itself. 
This viewpoint allows to find an implicit expression for the Painlev\'e transcendent in terms of Nekrasov-Okounkov partition functions as in 
eq.\eqref{eq:solution_isomonodromic}. In the autonomous limit, this reduces to an implicit form of the solution of the Calogero-Moser dynamics as studied in 
\cite{1999JMP....40.6339G}. As discussed in detail in Section \ref{sec:AutLimit},
this corresponds on the gauge theory side to the limit to the Seiberg-Witten geometry
and induces the explicit relation between the UV and IR gauge couplings of ${\mathcal N}=2^*$ theory as in equation \eqref{eq:SWUVCoupling}. In other words, Calogero-Moser dynamics describes the renormalisation group flow of ${\mathcal N}=2^*$  Seiberg-Witten theory. Turning on a self-dual $\Omega$-background amounts to 
move to the isomonodromic problem, that is the deautonomization of Calogero-Moser
and eq.\eqref{eq:solution_isomonodromic} describes the exact renormalisation group flow of the deformed theory.
Indeed, the isomonodromic time dependence includes the deformation of the SW curve in presence of a self-dual $\Omega$-background $\hbar$, in line with the expectation
of Painlev\'e/gauge theory correspondence.  This deformation accounts for all-genus topological string amplitudes on local Calabi-Yau geometries build on the relevant Seiberg-Witten curves \footnote{Similar results were recently obtained in the case of $SU(2)$ $N_f=4$ superconformal gauge theory in \cite{Coman:2018uwk}.}.

There is a number of open questions it would be worth to explore.
Let us observe that renormalisation group equations of the Seiberg-Witten theory 
can be obtained from blow-up equations as shown in detail in \cite{Nakajima:2003pg}
for the ${\mathcal N}=2$ SYM case. This approach should be generalised to the 
${\mathcal N}=2^*$ theory, the corresponding Painlev\'e equation arising from the 
relevant blowup equation.
The generalisation to higher rank and number of punctures will be presented in a separate paper. 
Generalization of the correspondence between isomonodromic
deformation problems on higher genus Riemann surfaces and the renormalization group flows of the corresponding class $S$ theory
would open new possibilities to investigate on one side the partition function of gauge theories with trifundamentals, and on the other give explicit constructions for the relevant tau-functions.

Another interesting venue is the inclusion of BPS observables in the gauge theory in the $\Omega$-background. Their couplings
can be regarded as additional time variables generating a hierarchy of extra flows which should provide the full character of $\mathcal{W}_{1+\infty}$
algebra.

\qquad
\qquad

{\bf Acknowledgements}\,\,\,\,
We would like to thank M.~Bershtein, H.~Desiraju, A.~Grassi, I.~Krichever, K.~Maruyoshi and F.~Morales for their interest in this work and for the numerous discussions. We would also like to thank the organizers of the workshops "Supersymmetric Quantum Field Theories in the Non-perturbative Regime" at GGI, Florence, and "Tau Functions of Integrable Systems and Their Applications" at BIRS, Banff, where part of the work was done.
The work of GB and FDM is supported by INFN via Iniziativa Specifica ST\&FI.
The work of PG was carried out within the HSE University Basic Research Program and funded by the Russian Academic Excellence Project '5-100'. The results of Section \ref{3} were obtained under the support of  Russian science foundation within the  grant 19-11-00275.  PG is also a \emph{Young Russian Mathematics} award winner and would like to thank its sponsors and jury.
The work of AT is supported by INFN via Iniziativa Specifica GAST and PRIN project "Geometria delle variet\'a algebriche".
The work of GB is supported by the PRIN project "Non-perturbative Aspects Of Gauge Theories And Strings". 

\vfill

\begin{appendix}

\section{Elliptic and theta functions}\label{sec:Elliptic}

For elliptic and theta functions we use the notations of \cite{DiFrancesco:1997nk}. Our torus has periods $(1,\tau) $, and the theta function that we use are
\eq{
\begin{split}
&\theta_1(z|\tau)\equiv -i\sum_{n\in\mathbb{Z}}(-1)^n q^{(n+\frac12)^2/2} e^{2\pi iz(n+\frac12)},\\
&\theta_2(z|\tau)\equiv\sum_{n\in\mathbb{Z}} q^{(n+\frac12)^2/2} e^{2\pi iz(n+\frac12)},\\
&\theta_3(z|\tau)\equiv \sum_{n\in\mathbb{Z}} q^{n^2/2} e^{2\pi iz n},\\
&\theta_4(z|\tau)\equiv \sum_{n\in\mathbb{Z}}(-1)^n q^{n^2/2} e^{2\pi izn},
\end{split}
}
where
\eq{
q=e^{2\pi i\tau}.
}
A prime denotes a derivative with respect to $z$, and when the theta function or its derivatives are evaluated at $z=0$, we simply denote it by $\theta_\nu(\tau)$ or $\theta_1'(\tau)$, e.t.c. Transformations of $\theta_1$ under elliptic transformations are
\begin{align}\label{eq:ThetaPeriod}
\theta_1(z+1|\tau)=-\theta_1(z|\tau), && \theta_1(z+\tau|\tau)=-q^{-1}e^{-2\pi i z}\theta_1(z|\tau).
\end{align}

In the main text we use also Weierstrass $\wp$ and $\zeta$. $\wp$ is a doubly periodic functions with a single double pole at $z=0$, that can be written in terms of $\theta_1$ as
\begin{equation}
\wp(z|\tau)=-\partial_z^2\log\theta_1(z|\tau)-2\eta_1(\tau)=\zeta'(z|\tau),
\end{equation}
where
\begin{equation}
\eta_1(\tau)=-\frac{1}{6}\frac{\theta_1'''(\tau)}{\theta_1'(\tau)}.
\end{equation}
Weierstrass' $\zeta$ function is minus the primitive of $\wp$. It has only one simple pole at $z=0$, and is quasi-elliptic:
\begin{equation}
\zeta(z|\tau)= 2\eta_1(\tau)z+\partial_z\log\theta_1(z|\tau),
\end{equation}
\begin{align}
\zeta(z+1|\tau)=\zeta(z|\tau)+ 2\eta_1(\tau), && \zeta(z+\tau|\tau)=\zeta(z|\tau)+ 2\tau\eta_1(\tau)-2\pi i.
\end{align}
Finally, we use Dedekind's $\eta$ function, defined as
\begin{equation}
\eta(\tau)=q^{1/24}\prod_{n=1}^\infty(1-q^n).
\end{equation}
It is related to the function $\theta_1$ by
\begin{equation}
\eta(\tau)=\left(\frac{\theta_1'(\tau)}{2\pi} \right)^{1/3}.
\end{equation}

Because of the periodicities \eqref{eq:ThetaPeriod}, the elliptic transformations of the Lam\'e function
\begin{equation}
x(u,z)=\frac{\theta_1(z-u|\tau)\theta_1'(\tau)}{\theta_1(z|\tau)\theta_1(u|\tau)}
\end{equation}
are given by
\begin{align}
x(u,z+1)=x(u,z), && x(u,z+\tau)=e^{2\pi i u}x(u,z).
\end{align}
The product of Lam\'e functions satisfies the following identities:
\begin{align}\label{eq:LameWP}
x(u,z)x(-u,z)=\wp(z)-\wp(u), && x(u,z)y(-u,z)-y(u,z)x(-u,z)=\wp'(u),
\end{align}
where $y(u,z)=\partial_ux(u,z)$, that are used in computing the isomonodromic Hamiltonian $H_\tau$ from the Lax matrix. Further, to show that the zero-curvature equation \eqref{eq:ZeroCurv} is the compatibility condition for the system \eqref{eq:LaxPairSystem}, one has to use the property
\begin{equation}\label{eq:LameProp0}
2\pi i\partial_\tau x(u,z)+\partial_z\partial_u x(u,z)=0.
\end{equation}
The following theta-function identities are used in the study of the autonomous limit:
\eq{\label{eq:ThetaId1}
\pd_z\frac{\theta_2(z|\tau)}{\theta_3(z|\tau)}=-\pi\theta_4^2(\tau)\frac{\theta_1(z|\tau)\theta_4(z|\tau)}{\theta_3(z|\tau)^2},
}
\eq{\label{eq:ThetaId2}
\frac{\theta_1(2Q|2\tau_0)^2}{\theta_3(2Q|2\tau_0)^2}=
\frac{\theta_2(2\tau_0)^2}{\theta_4(2\tau_0)^2}-\frac{\theta_3(2\tau_0)^2}{\theta_4(2\tau_0)^2}\frac{\theta_2(2Q|2\tau_0)^2}{\theta_3(2Q|2\tau_0)^2},\\
\frac{\theta_4(2Q|2\tau_0)^2}{\theta_3(2Q|2\tau_0)^2}=
\frac{\theta_3(2\tau_0)^2}{\theta_4(2\tau_0)^2}-\frac{\theta_2(2\tau_0)^2}{\theta_4(2\tau_0)^2}\frac{\theta_2(2Q|2\tau_0)^2}{\theta_3(2Q|2\tau_0)^2}.
}

\section{Fusion and braiding of degenerate fields}\label{sec:Fusion}

In this paper monodromies of the fundamental solution have been computed by braiding and fusion of degenerate fields with Virasoro primaries. Fusion matrices linearly relate different sets of blocks, and analytic continuation of a conformal block along any arbitrary contour can be decomposed in elementary braiding and fusion rules \cite{moore1989}. 

In this paper we only need the following braiding move,
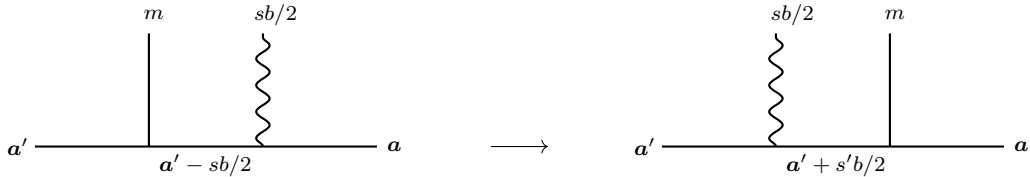
\begin{figure}[h!]
\begin{tikzpicture}\label{fig:BraidingMove}
\tikzmath{\scale=1.5;\radius=1;\offset=1/3;}
\begin{scope}[scale=\scale]
\draw[thick](0,0)--(\radius,0);
\draw[thick](\radius,0)--(\radius,\radius);
\draw[thick](\radius,0)--(2*\radius,0);
\draw[thick](2*\radius,0)--(3*\radius,0);
\draw[thick,wave](2*\radius,0)--(2*\radius,\radius);
\node[anchor=west] at(\radius-\offset/2.5,\radius+\offset/2) {\scriptsize $ m$};
\node[anchor=west] at(2*\radius-\offset/2,\radius+\offset/2) {\scriptsize $sb/2$};
\node[anchor=west] at(\radius,-\offset/2) {\scriptsize $\bs a'-sb/2 $};
\node[anchor=west] at(-\radius+2*\offset,0) {\scriptsize $\bs a'$};
\node[anchor=west] at(3*\radius,0) {\scriptsize $\bs a$};
\draw[->](4,0)--(4.5,0);
\end{scope}

\begin{scope}[scale=\scale,xshift=5.5cm]
\draw[thick](0,0)--(\radius,0);
\draw[thick,wave](\radius,0)--(\radius,\radius);
\draw[thick](\radius,0)--(2*\radius,0);
\draw[thick](2*\radius,0)--(3*\radius,0);
\draw[thick](2*\radius,0)--(2*\radius,\radius);
\node[anchor=west] at(\radius-\offset/2.5,\radius+\offset/2) {\scriptsize $sb/2$};
\node[anchor=west] at(-\radius+2*\offset,0) {\scriptsize $\bs a'$};
\node[anchor=west] at(2*\radius-\offset/2.5,\radius+\offset/2) {\scriptsize $m$};
\node[anchor=west] at(3*\radius,0) {\scriptsize $\bs a$};
\node[anchor=west] at(\radius,-\offset/2) {\scriptsize $\bs a'+s'b/2 $};
\end{scope}
\end{tikzpicture}
\caption{Braiding of a degenerate field and a primary}
\end{figure}

that correspond to the following analytic continuation operation on the matrix elements\footnote{comparing to \cite{Gavrylenko:2018ckn} we changed here the sign of the fusion matrix, so that for $m=0$ braiding is trivial}:
\begin{equation}
\langle \bs a' |V_{\bs m}(y) \phi(\gamma_y\cdot z)|\bs a\rangle=i\sum_{s'=\pm} \langle \bs a' |\phi_{s'}(z) V_m(y) |\bs a\rangle e^{i\pi s'a'}F_{s's}(a',m,a)e^{-i\pi sa},
\end{equation}
where $\gamma_y\cdot z$ is the analytic continuation of the degenerate field along a contour going around $y$ in the positive direction, on the sphere from zero to infinity. $F$ is the fusion matrix, that with our normalization conventions takes the form
\begin{equation}\label{eq:FusionMatrix}
F(a',m,a)=\left( \begin{array}{cc}
\frac{\cos\pi(m+a+a')}{\sin2\pi a} & \frac{\cos\pi(m+a'-a)}{\sin 2\pi a} \\ -\frac{\cos\pi(m+a-a')}{\sin2\pi a} & -\frac{\cos\pi(m-a-a')}{\sin2\pi a}
\end{array} \right).
\end{equation}

\section{Periodicity of the tau functions}\label{sec:Periodicity}

Equation \eqref{eq:deaut_Calogero} is invariant under the transformation $Q\mapsto Q+\frac n2 +\frac{\tau k}2$, where $k,n\in\mathbb Z$.
We denote this transformation by $\delta_{\frac n2,\frac k2}$.
One might ask the following question: what are the transformation properties of the tau function and dual partition functions under $\delta_{\frac n2,\frac k2}$?

The tau function after the transformation is defined by
\eq{
2\pi i\pd_\tau \log(\delta_{\frac n2,\frac k2}{\mc T})=(2\pi i)^2(\pd_\tau Q+k/2)^2-m^2\wp(2Q)-2m^2\eta_1(\tau)=\\
=2\pi i\pd_\tau {\mc T}+(2\pi i)^2(k^2/4+k\pd_\tau Q).
}
Therefore
\eq{
\delta_{\frac n2,\frac k2}{\mc T}=C_{\frac n2,\frac k2}\cdot q^{k^2/4}e^{2\pi i kQ}\mc T
}
Using now \eqref{eq:ZD-tau} we compute transformations of $Z_0^D$, $Z_{1/2}^D$:
\eq{
\delta_{\frac n2,\frac k2}Z_{0}^D=C_{\frac n2,\frac k2}\eta(\tau)^{-1}\theta_3(2Q+n+k\tau|2\tau)q^{k^2/4}e^{2\pi ikQ}\mc T=\\=
\left[\begin{array}{l}
k\in 2\mathbb Z: C_{\frac n2,\frac k2}\eta(\tau)^{-1}\theta_3(2Q\tau|2\tau)\mc T=C_{\frac n2,\frac k2}Z_{0}^D\\
k\in 1+2\mathbb Z: C_{\frac n2,\frac k2}\eta(\tau)^{-1}\theta_2(2Q\tau|2\tau)\mc T=C_{\frac n2,\frac k2}Z_{1/2}^D
\end{array}
\right.
}
In this way we see that the dual partition functions have much better behaviour than the tau function $\mc T$.
As we will see later, such shifts of parameters correspond to shifts of the initial data:
\eq{
\delta_{\frac n2,\frac k2}(\eta,a)=(\eta+2\pi n,a+\frac k2)
}
Therefore transformations of the dual partition functions look as follows:
\eq{
\delta_{\frac n2,\frac k2}Z^D_{1/2}=e^{-i\eta \frac k2}\cdot
\left[\begin{array}{l}
k\in 2\mathbb Z: Z^D_{1/2}\\
k\in 1+2\mathbb Z:  Z^D_{0}
\end{array}
\right.
}
Therefore we can conclude that $C_{\frac n2,\frac k2}=e^{-i\eta\frac k2}$, so $\delta_{\frac n2,\frac k2}\mc T=q^{k^2/4}e^{ik(2\pi Q-\frac\eta 2)}\mc T$.

\section{Asymptotic calculation of the tau function}\label{eq:AppNum}

\subsection{Algorithm of computations}

We start from the following ansatz for the solution of the non-autonomous Calogero equation:

\eq{\label{eq:Q_expansion}
Q=\alpha\tau+\beta+\frac1{2\pi i}\sum_{n=0}^{\infty}\sum_{k=-n}^\infty c_{n,k} q^n e^{4\pi i k(\alpha\tau+\beta)}=\alpha\tau+\beta + \frac1{2\pi i}X
}

Series expansion of $\wp(z|\tau)+2\eta_1(\tau)=-\pd_z^2\log\theta_1(z|\tau)$ looks as follows:
\eq{
\frac1{(2\pi i)^2}(-\pd_z^2\log\theta_1(z|\tau))=\sum_{n=0}^\infty\sum_{k=1}^\infty k q^{kn}e^{2\pi ikz}+
\sum_{n=1}^\infty\sum_{k=1}^\infty k q^{kn}e^{-2\pi ikz}
}
Rewrite now equation \eqref{eq:deaut_Calogero} using last two formulas and introducing notation $s=e^{2\pi i(\alpha\tau+\beta)}$.
We also introduce single formal parameter of expansion $\epsilon$ in the following way: $q\mapsto q\cdot\epsilon^2$, $s\mapsto s\cdot\sqrt{\epsilon}$.
\eq{
\sum_{n=0}^\infty\sum_{k=-n}^\infty (n+2\alpha k)^2 c_{n,k} q^n s^{2k}\epsilon^{2n+k}=\\=
m^2\sum_{n=0}^\infty\sum_{k=1}^\infty k^2 q^{kn} s^{2k} e^{k X(q,s)}\epsilon^{k(2n+1)}-
m^2\sum_{n=1}^\infty\sum_{k=1}^\infty k^2 q^{kn} s^{-2k}e^{-kX(q,s)}\epsilon^{k(2n-1)}
}
One can see that powers of $\epsilon$ in the r.h.s. are at least one, therefore higher-order coefficients $c_{n,k}$ become functions of the lower-order ones, and thus equation can be solved order-by-order starting from $c_{0,0}=0$~\footnote{This algorithm is not optimal for the computation of non-trivial coefficients: as we will see later, it gives a lot of zero terms in the dual partition functions.}.

After this is done, we need to compute the logarithm of the isomonodromic tau function:
\eq{
\log\mc T=\alpha^2\log q + \sum_{n,k} q^n s^k b_{n,k}=\alpha^2\log q+Y(q,s)\\
\sum_{n,k}(n+2\alpha k)q^ns^kb_{n,k}
=\left(\sum_{n,k}(n+2\alpha k)q^ns^{2k}c_{n,k}\right)^2-\\
-m^2\left(\sum_{n=0}^\infty\sum_{k=1}^\infty k q^{kn}s^{2k} e^{k X(q,s)}+
\sum_{n=1}^\infty\sum_{k=1}^\infty k q^{kn}s^{-2k} e^{-k X(q,s)}
\right)
}
Then we compute the two dual partition functions:
\eq{
Z^D_{0}=q^{\alpha^2-1/24}\sum_{n=1}^\infty p(n)q^n\cdot \sum_{n=-\infty}^\infty q^{n^2}s^{2n}e^{2nX(q,s)}\cdot e^{Y(q,s)+2\alpha X(q,s)}=\sum_{n,k}z_{n,k}q^{n+\alpha^2-1/{24}} s^{2k},\\
Z^D_{1/2}=q^{\alpha^2-1/24}\sum_{n=1}^\infty p(n)q^n\cdot \sum_{n=-\infty}^\infty q^{(n+\frac12)^2}s^{2n+1}e^{(2n+1)X(q,s)}\cdot e^{Y(q,s)+2\alpha X(q,s)}=\\=\sum_{n,k}z'_{n,k}q^{n+\alpha^2+5/{24}} s^{2k-1}.
}

\subsection{Results}

We solved the equation asymptotically up to $\epsilon^6$.

Important information about the function $f(q,s)=\sum f_{n,k}q^n s^{2k}$ is encoded in the list of non-zero coefficients $f_{n,k}$:
we will denote such coefficients by points $(k,n)$ in the integer plane (sometimes it will be shifted by some fractional numbers which we neglect). We call the set of such points \emph{support} of the function $f$. 

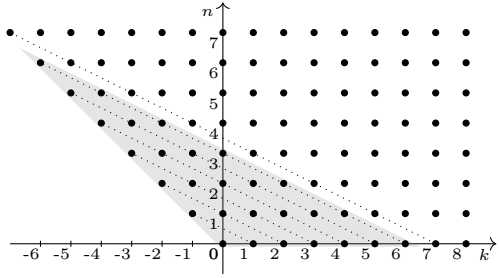
\begin{figure}[h]
\begin{center}
\begin{tikzpicture}[scale=0.4,font=\tiny]

\draw[->](0,-1)-- (0,8);
\draw[->](-7,0)--(9,0);
\path[fill=gray,opacity=0.2](6.5,-0.1)--(-6.7,6.5)--(-0.1,-0.1);
\foreach \j in {0,...,7}
\foreach \i in {-8,...,\j} \draw[fill=black] (\i-\j+8,\j) circle(0.1);

\foreach \i in {0,...,7} \draw[dotted](\i,0)--(-\i,\i);

\foreach \i in {-6,...,8}{ \node at (\i-0.3,-0.35){\i}; \draw(\i,-0.1)--(\i,0.1);};

\foreach \j in {0,...,7} {\node at (-0.3,\j-0.35){\j};};

\node at (-0.5,7.65){$n$};
\node at (8.6,-0.35){$k$};

\end{tikzpicture}
\end{center}
\caption{Support of $X(q,s)$}
\label{fig:support1}
\end{figure}

First we show the support of the solution $X(q,s)$ in Fig.~\ref{fig:support1}. In this picture the gray region contains all  coefficients that were computed in the asymptotic expansion.
Dotted lines show monomials with the same order of $\epsilon$. The full support is bounded from below by the lines $n=0$ and $n=-k$.

The first few terms of the expansion look as follows:
\eq{
c_{0,1}=\frac{m^2}{4 \alpha ^2},\quad c_{1,-1}=-\frac{m^2}{(2 \alpha-1)^2},\quad
c_{0,2}=\frac{m^2 \left(8 \alpha ^2+m^2\right)}{32 \alpha ^4},\quad\\ c_{1,0}= -\frac{(4 \alpha -1) m^4}{2 \alpha ^2 (2 \alpha -1)^2},\quad
c_{2,-2}=-\frac{m^2\left(8 \alpha ^2-8 \alpha +m^2+2\right)}{2 (2 \alpha -1)^4},\quad \ldots
}

\begin{figure}[h]
\begin{center}
\begin{tabular}{cc}

\begin{tikzpicture}[scale=0.4,font=\tiny]

\draw[->](0,-1)-- (0,8);
\draw[->](-7,0)--(9,0);
\path[fill=gray,opacity=0.2](6.5,-0.1)--(-6.7,6.5)--(-0.1,-0.1);
\foreach \i [evaluate=\i as \x using \i*\i] in {-2,...,2}
\foreach \j  in {\x,...,7} \draw[fill=black] (\i,\j) circle(0.1);


\foreach \j in {0,...,7}
\foreach \i in {-8,...,\j} \draw[fill=black] (\i-\j+8,\j) circle(0.02);

\foreach \i in {-6,...,8}{ \node at (\i-0.3,-0.35){\i}; \draw(\i,-0.1)--(\i,0.1);};
\foreach \j in {0,...,7} {\node at (-0.3,\j-0.35){\j};};

\draw [domain=-2.65:2.65,dashed] plot ({\x},{\x*\x});

\node at (-0.5,7.65){$n$};
\node at (8.6,-0.35){$k$};

\end{tikzpicture}

&

\begin{tikzpicture}[scale=0.4,font=\tiny]

\draw[->](0,-1)-- (0,8);
\draw[->](-7,0)--(9,0);
\path[fill=gray,opacity=0.2](6.5,-0.1)--(-6.7,6.5)--(-0.1,-0.1);
\foreach \i [evaluate=\i as \x using \i*\i-\i] in {-2,...,3}
\foreach \j  in {\x,...,7} \draw[fill=black] (\i,\j) circle(0.1);

\foreach \j in {0,...,7}
\foreach \i in {-8,...,\j} \draw[fill=black] (\i-\j+8,\j) circle(0.02);

\foreach \i in {-6,...,8}{ \node at (\i-0.3,-0.35){\i}; \draw(\i,-0.1)--(\i,0.1);};
\foreach \j in {0,...,7} {\node at (-0.3,\j-0.35){\j};};

\node at (-0.5,7.65){$n$};
\node at (8.6,-0.35){$k$};

\draw [domain=-2.25:3.25,dashed] plot ({\x},{\x*\x-\x});

\end{tikzpicture}

\end{tabular}

\end{center}
\caption{Support of $Z_{0}^D$ (left) and of $Z_{1/2}^D$ (right). }
\label{fig:support2}
\end{figure}
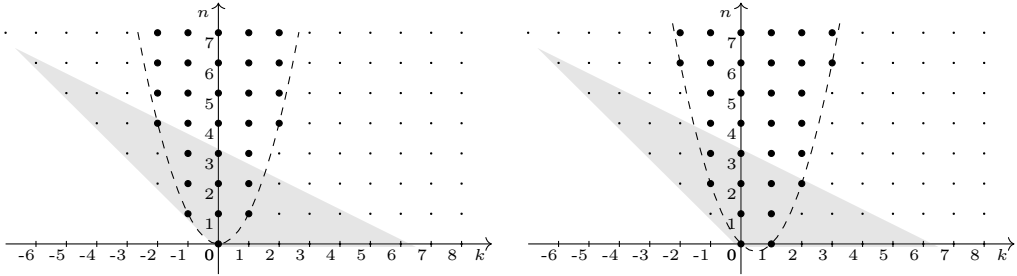

The support of the dual partition functions is shown in Fig.~\ref{fig:support2}.

We see that some non-trivial cancellation happened and a lot of coefficients that naively might be non-zero (denoted by small dots) actually vanish.

Values of the first non-trivial coefficients are given by

\eq{
z_{0,0}=1,\quad z'_{0,0}=1,\quad z'_{0,1}=1-\frac{m^2}{4\alpha^2},\quad
z_{1,-1}=\frac{ (m^2-(1-2 \alpha ))^2}{(1-2 \alpha )^2},\\
z_{1,1}=\frac{(m^2-4 \alpha ^2)^2 (m^2-(2 \alpha
   +1)^2)}{16 \alpha ^4 (2 \alpha +1)^2},\quad \ldots
}

We also found experimentally that normalized values of all other non-trivial coefficients can be given in terms of a single function, which will be identified with the toric conformal block:
\eq{
\mc B(a,m,q)=1+q \left(\frac{(m^2-1) m^2}{2 a^2}+1\right)+q^2 \left(\frac{3
   (m^2-4) (m^2-1)^2 m^2}{16
   (a^2-\frac{1}{4})^2}-\right.\\\left.-\frac{(m^2-3) (m^2-1)
   (m^2+2) m^2}{4 (a^2-\frac{1}{4})}+\frac{(m^2-1)
   (m^4-m^2+2) m^2}{4 a^2}+2\right)+\\+q^3 \left(\frac{(m^2-9)
   (m^2-4)^2 (m^2-1)^2 m^2}{36
   (a^2-1)^2}+\frac{(m^2-4) (m^2-1)^2 (2 m^4-2
   m^2+9) m^2}{48 (a^2-\frac{1}{4})^2}-\right.\\-\frac{(m^2-4)
   (m^2-1) (11 m^6-106 m^4+131 m^2+108) m^2}{216
   (a^2-1)}+\\+\frac{(m^2-1) (3 m^8-22 m^6+65 m^4-46
   m^2+24) m^2}{24 a^2}-\\\left.-\frac{(m^2-1) (8 m^8-24 m^6+15
   m^4+m^2-162) m^2}{108 (a^2-\frac{1}{4})}+3\right)+O(q^3)=\sum_{n=0}^\infty \mc B_n(a,m)q^n
}
Namely, the ratios of coefficients are
\eq{
z_{n,0}/z_{0,0}=\mc B_n(\alpha,m),\quad n=0,1,2,3,\quad
z_{2,1}/z_{1,1}=\mc B_1(\alpha+1,m),\quad n=0,1,2,\\
\frac{z_{n+1,-1}}{z_{1,-1}}=\mc B_n(\alpha-1,m),\quad
\frac{z_{3,1}}{z_{1,1}}=\mc B_2(\alpha-1,m),\quad \frac{z'_{-1,3}}{z'_{-1,2}}=\mc B_1(a-\frac32,m)\\
\frac{z'_{n,0}}{z'_{0,0}}=\mc B_n(\alpha-\frac12,m),\quad n=0,1,2,3,\quad
\frac{z'_{n,1}}{z'_{0,1}}=\mc B_n(\alpha+\frac12,m),\quad n=0,1,2.\quad
}
We see that the latter formula is in complete agreement with \eqref{eq:TauFuncVirasoro} if $\mc B$ is a conformal block, so we check that it actually coincides with the AGT formula
\eq{
\label{eq:conf_block_AGT}
\mc B(a,m,q)=\prod_{n=1}^\infty(1-q^n)^{1-2m^2}\sum_{Y_+,Y_-}q^{|Y_+|+|Y_-|}\prod_{\epsilon,\epsilon'=\pm}\frac{N_{Y_\epsilon,Y_{\epsilon'}}(m+(\epsilon-\epsilon') a)}{N_{Y_\epsilon,Y_{\epsilon'}}((\epsilon-\epsilon') a)}
}
where Nekrasov factors are given by
\eq{
N_{\lambda,\mu}(x)=\prod_{s\in\lambda}(x+a_\lambda(s)+l_\mu(s)+1)
\prod_{t\in\mu}(x-l_\lambda(t)-a_\mu(t)-1).
}

\subsection{Asymptotic computation of monodromies}

At the moment we have two different parameterization: the one in terms of $(\alpha,\beta)$, initial data of the equation, and the other one in terms of the monodromy data $(a,\eta)$.
We need to know the explicit identification between them. To compute this identification we use the fact that the evolution is isomonodromic, so monodromies can be computed in the limit $\tau\to+i\infty$.

\eq{
\theta_1(z|\tau)=2q^{1/8}\left(\sin\pi z-q \sin 3\pi z+\ldots\right).
}
One can take just the first term of expansion until it becomes smaller than the first correction.
This occurs when $\sin\pi z\approx e^{2 \pi i\tau}\sin{3\pi z}$, so for $z=\pm\tau$. We will choose two copies of the A-cycle with ${\rm Im} z=\pm {\rm Im} \tau/2$ and work in the region between them.
So for all computations to be consistent we need to have $-1/4<{\rm Re}\alpha<1/4$ (one can easily overcome this constraint taking more terms of expansion).
Also convergence of the series \eqref{eq:Q_expansion} requires $\alpha>0$. So for simplicity we just take $\alpha$ to have a sufficiently small positive real part.

Our approximation for $x$ is then
\eq{
x(u,z)\approx\frac{\pi\sin\pi(z-u)}{\sin\pi z\sin\pi u}
=\frac{2\pi i}{e^{2\pi iu}-1}-\frac{2\pi i}{e^{2\pi iz}-1}.
}
The first terms of the expansion of $Q$ look as
\eq{
Q(\tau)=\alpha\tau+\beta+\frac{m^2}{8\pi i\alpha^2}e^{4\pi i(\alpha\tau+\beta)}+\ldots
}
The leading behavior of the connection matrix is
\eq{
L(z|\tau)=2\pi i\begin{pmatrix}
\alpha+\frac{m^2}{2\alpha}e^{4\pi i(\alpha\tau+\beta)}&
\frac{m}{e^{4\pi i(\alpha\tau+\beta)}-1}-\frac{m}{e^{2\pi iz}-1}\\
\frac{m}{e^{-4\pi i(\alpha\tau+\beta)}-1}-\frac{m}{e^{2\pi iz}-1}&
-\alpha-\frac{m^2}{2\alpha}e^{4\pi i(\alpha\tau+\beta)}\end{pmatrix}.
}
Further expanding up to first order in $e^{4\pi i\alpha\tau}$ we get
\eq{
L(z|\tau)=2\pi i\begin{pmatrix}
\alpha&-\frac{m e^{2\pi i z}}{e^{2\pi iz}-1}\\
-\frac{m}{e^{2\pi iz}-1}&-\alpha\end{pmatrix}+
2\pi i e^{4\pi i(\alpha\tau+\beta)}\begin{pmatrix}
\frac{m^2}{2\alpha}&-m\\ m& -\frac{m^2}{2\alpha}\end{pmatrix}.
}
One may notice that there is an equality
\eq{
L(z|\tau)=U_0L_0(z|\tau)U_0^{-1},
}
where $U_0=1+\frac{m}{2\alpha}e^{4\pi i(\alpha\tau+\beta)}\sigma^x$. This equality is a reminiscence of the isomonodromic deformation equation \eqref{eq:LaxPairSystem}.
The solution of the linear system in the region $z\sim\frac\tau2$ is given by
\eq{\label{eq:AsymptoticSol}
Y(z)=(1-e^{2\pi iz})^mU_0\times\\\times
\begin{pmatrix}
\Fhyp(m,m+2\alpha,2\alpha,e^{2\pi iz})&
\frac{me^{2\pi i z}}{1-2\alpha}\Fhyp(1+m,1+m-2\alpha,2-2\alpha,e^{2\pi iz})\\
\frac{m}{2\alpha}\Fhyp(1+m,m+2\alpha,1+2\alpha,e^{2\pi iz})&
\Fhyp(m,1+m-2\alpha,1-2\alpha,e^{2\pi i z})
\end{pmatrix}\times\\\times
{\rm diag}((-e^{2\pi i z})^\alpha,(-e^{2\pi i z})^{-\alpha})
}
Now we compute the analytic continuation of this solution to the region $z\sim-\frac\tau 2$ along imaginary line ${\rm Re} z=1/2$~\footnote{To do this we use connection formula for hypergeometric function
\eq{
\Fhyp(a,b,c,z)=\frac{\Gamma(c)\Gamma(b-a)}{\Gamma(b)\Gamma(c-a)}(-z)^{-a} \Fhyp(a,a-c+1,a-b+1,z^{-1})+\\+
\frac{\Gamma(c)\Gamma(a-b)}{\Gamma(a)\Gamma(c-b)}(-z)^{-b}\Fhyp(b,b-c+1,b-a+1,z^{-1})
}
}
:
\eq{\label{eq:MB_tilde}
Y(z)=\sigma^x Y(-z) \sigma^x \widetilde{M_B},\qquad \widetilde{M_B}=
\begin{pmatrix}
 \frac{\Gamma(2\alpha)^2}{\Gamma(2\alpha+m)\Gamma(2\alpha-m)}&
\frac{\Gamma(2-2\alpha)\Gamma(-1+2\alpha)}{\Gamma(m)\Gamma(1-m)}\\
 \frac{\Gamma(1-2\alpha)\Gamma(2\alpha)}{\Gamma(1-m)\Gamma(m)}&
\frac{\Gamma(1-2\alpha)^2}{\Gamma(1-m-2\alpha)\Gamma(1+m-2\alpha)}
\end{pmatrix}.
}
We wish to compute B-cycle monodromy. Its defining relation is
\eq{
Y(\frac\tau2+x)=e^{2\pi i\bs Q}Y(-\frac\tau2+x)M_B.
}
Using \eqref{eq:MB_tilde} we get
\eq{
Y(\frac\tau2+x)=e^{2\pi i\bs Q}\sigma^x Y(\frac\tau2-x)\sigma^x \widetilde{M_B}M_B.
}
We write this expression down keeping only the first orders:
\eq{
(1+\frac m{2\alpha}e^{4\pi i(\alpha\tau+\beta)}\sigma^x)
\begin{pmatrix}
1&0\\ \frac{m}{2\alpha}&1
\end{pmatrix}
\times\diag(e^{\pi i\alpha \tau},e^{-\pi i\alpha\tau})
\times\diag((-e^{2\pi i x})^{\alpha},(-e^{2\pi ix})^{-\alpha})=\\=
\diag\left(e^{2\pi i(\alpha\tau+\beta)},e^{-2\pi i(\alpha\tau+\beta)}+\frac{m^2}{4\alpha^2}e^{2\pi i(\alpha\tau+\beta)}\right)
(1+\frac m{2\alpha}e^{4\pi i(\alpha\tau+\beta)}\sigma^x)
\begin{pmatrix}
1&\frac{m}{2\alpha}\\0&1
\end{pmatrix}
\times\\\times
\diag(e^{-\pi i\alpha \tau},e^{\pi i\alpha\tau})
\times\diag((-e^{2\pi i x})^{\alpha},(-e^{2\pi ix})^{-\alpha})\times \widetilde{M_B} M_B.
}
By using also the relation
\begin{equation}
\sigma^x \widetilde{M_B} \sigma^x = \widetilde{M_B}^{-1}
\end{equation}
we can then express $M_B$ as
\eq{
M_B=\sigma^x\widetilde{M_B}\sigma^x\times\diag(e^{-2\pi i\beta},e^{2\pi i\beta})+O(e^{4\pi i\alpha\tau}).
}
We see that to our precision $M_B$ is actually constant when $\tau\to i\infty$, its value is given by
\eq{
M_B=
\begin{pmatrix}
\frac{\Gamma(1-2\alpha)^2}{\Gamma(1-m-2\alpha)\Gamma(1+m-2\alpha)}e^{-2\pi i\beta}&
\frac{\sin\pi m}{\sin\pi\alpha}e^{2\pi i\beta}\\
-\frac{\sin\pi m}{\sin\pi\alpha}e^{-2\pi i\beta}&
 \frac{\Gamma(2\alpha)^2}{\Gamma(2\alpha+m)\Gamma(2\alpha-m)}e^{2\pi i\beta}
\end{pmatrix}.
}
To compare this with the monodromy matrix \eqref{eq:TorusMonodromies} computed by braiding we now change normalization
\begin{align}
e^{2\pi i\beta}=e^{2\pi i\beta'}/r, && M_B'=\diag(r^{1/2},r^{-1/2}) M_B \diag(r^{-1/2},r^{1/2}),
\end{align}
where
\begin{equation}
r=\frac{\Gamma(2\alpha)\Gamma(1-2\alpha-m)}{\Gamma(1-2\alpha)\Gamma(2\alpha-m)}.
\end{equation}
The new monodromy is
\eq{
\label{eq:MB_num}
M_B'=\begin{pmatrix}
\frac{\sin \pi(2\alpha-m)}{\sin 2\pi\alpha}e^{-2\pi i\beta'}&
\frac{\sin\pi m}{\sin2\pi\alpha}e^{2\pi i\beta'}\\
-\frac{\sin\pi m}{\sin2\pi\alpha}e^{-2\pi i\beta'}&
\frac{\sin\pi(2\alpha+m)}{\sin2\pi\alpha}e^{2\pi i\beta'}
\end{pmatrix}.
}
Corresponding A-cycle monodromy is clearly given by the formula
\eq{
\label{eq:MA_num}
M_A'=M_A=\begin{pmatrix}e^{2\pi i a}&0\\0&e^{-2\pi ia}\end{pmatrix}.
}


These monodromies are related to those computed from CFT \eqref{eq:TorusMonodromies} by a conjugation with the matrix
\begin{equation}
C=\begin{pmatrix}0&-i e^{\pi i a}\\i e^{-\pi i a}&\end{pmatrix}.
\end{equation}
We can in fact check explicitly that
\eq{
M_A'=CM_A^{(CFT)}C^{-1}=\begin{pmatrix}e^{2\pi ia}&0\\0&e^{-2\pi i a}
\end{pmatrix},\\
M_B'=CM_B^{(CFT)}C^{-1}=
\begin{pmatrix}
\frac{\sin \pi(2 a-m)}{\sin 2\pi a}e^{-i\eta/2}&
\frac{\sin\pi m}{\sin2\pi a}e^{i\eta/2}\\
-\frac{\sin\pi m}{\sin2\pi a}e^{-i\eta/2}&
\frac{\sin\pi(2 a+m)}{\sin2\pi a}e^{i\eta/2}
\end{pmatrix},
}
after we made the identification
\begin{align}
\alpha=a, && \eta=4\pi\beta'.
\end{align}

\subsection{Identification of parameters}

One may check that the two  asymptotic expansions have the following forms:
\eq{
\eta(\tau)^{-1}\theta_3(2Q|\tau)\mc T(\tau)=\sum_{n\in\mathbb Z} C_n e^{4\pi i n\beta} q^{(\alpha+n)^2} \mc B(\alpha+n,m,q)\\
\eta(\tau)^{-1}\theta_2(2Q|\tau)\mc T(\tau)=\sum_{n\in\mathbb Z+\frac12} C_n e^{4\pi i n\beta} q^{(\alpha+n)^2} \mc B(\alpha+n,m,q)\\
}
Where the structure constants are given by explicit formula
\eq{
C_n=\frac{G(1-m+2(\alpha+n))G(1-m-2(\alpha+n)}{G(1+2(\alpha+n))G(1-2(\alpha+n))}
\frac{G(1+2\alpha)G(1-2\alpha)}{G(1-m+2\alpha)G(1-m-2\alpha)}\\\times
\left(\frac{\Gamma(2\alpha)\Gamma(1-m-2\alpha)}{\Gamma(1-2\alpha)\Gamma(2\alpha-m)}\right)^{2n}.
}
We may check, in particular, that $C_0=C_{-\frac12}=1$, which is consistent with experimental results. We also see that after the redefinition\footnote{Note this is the same redefinition of \eqref{eq:BetaSec3}, if $\eta=4\pi\beta'$, which we will see is the case.}
\eq{
\label{eq:beta_appendix_1}
e^{2 i\beta'}=e^{2i\beta}\frac{\Gamma(2\alpha)\Gamma(1-m-2\alpha)}{\Gamma(1-2\alpha)\Gamma(2\alpha-m)}.
}
Experimentally found tau functions may be rewritten as
\eq{
\theta_3(2Q|\tau)\mc T(\tau)=C(\alpha)^{-1}\sum_{n\in\mathbb Z} C(\alpha+n) e^{4\pi i n\beta'} q^{(\alpha+n)^2} \mc B(\alpha+n,m,q)\\
\theta_2(2Q|\tau)\mc T(\tau)=C(\alpha)^{-1}\sum_{n\in\mathbb Z+\frac12} C(\alpha+n) e^{4\pi i n\beta'} q^{(\alpha+n)^2} \mc B(\alpha+n,m,q)\\
}
where
\eq{
C(\alpha)=\frac{G(1-m+2\alpha)G(1-m-2\alpha)}{G(1+2\alpha)G(1-2\alpha)}
}

Again, comparing with expressions \eqref{eq:ZD-tau} we find $a=\alpha$, $\eta=4\pi\beta'$, consistently with what we found in the asymptotic computation of the monodromy matrices.


\section{A self-consistency check}\label{sec:SelfConst}

Here we do some calculation to perform extra check of $\tau$-derivatives of some correlators. Namely, compute
\eq{
F(w,y)_{\alpha\beta}=2\pi i\pd_\tau\left(Y(w)^{-1}\Xi(y-w)Y(y)\right)_{\alpha\beta}=2\pi i\pd_\tau\frac{\langle\bar\psi_\alpha(w) \psi_\beta(y) V_m(0)\rangle}{\langle V_m(0)\rangle}=\\=
2\pi i\frac{\pd_\tau\langle\bar\psi_\alpha(w) \psi_\beta(y) V_m(0)\rangle}{\langle V_m(0)\rangle}-
2\pi i\frac{\langle\bar\psi_\alpha(w) \psi_\beta(y) V_m(0)\rangle}{\langle V_m(0)\rangle}
\frac{\pd_\tau\langle V_m(0)\rangle}{\langle V_m(0)\rangle}.
\label{eq:F_initial}
}
We then note that the derivative of any correlator with respect to the modular parameter $\tau$ can be realized by an integral over the A-cycle of an insertion of the energy-momentum tensor $T(z)$:
\begin{equation}
2\pi i\partial_\tau\langle\mathcal{O}\rangle=\oint_A\langle\mathcal{O}T(z)\rangle-\langle T\rangle\langle\mathcal{O}\rangle.
\end{equation}
Further, we can use the explicit expression of the free fermion energy-momentum tensor

\eq{
T(z)=\frac12\sum_\gamma\left(:\pd\bar\psi_\gamma(z)\psi_\gamma(z):+:\pd\psi_\gamma(z)\bar\psi_\gamma(z):\right)
}
where $:\ :$ denotes regular part of the OPE. Then, 
\eq{
F(w,y)_{\alpha\beta}=\sum_\gamma\oint_A dz \frac{\langle:\pd\bar\psi_\gamma(z)\psi_\gamma(z):\bar\psi_\alpha(w) \psi_\beta(y) V_m(0)\rangle
\langle V_m(0)\rangle}
{\langle V_m(0)\rangle^2}-\\-
\frac{\langle:\pd\bar\psi_\gamma(z)\psi_\gamma(z):V_m(0)\rangle\langle\bar\psi_\alpha(w) \psi_\beta(y) V_m(0)\rangle}{\langle V_m(0)\rangle^2}.
}
Here we added a total derivative to $T(z)$ since it does not change correlator.
Now we can compute this expression using the generalized Wick theorem~\cite{Alexandrov:2012tr}. The second term cancels with the term, containing pairing between $\bar\psi_\alpha(w)$ and $\psi_\beta(y)$. So finally we get only one term:
\eq{
F(w,y)_{\alpha\beta}=\sum_\gamma\oint_A dz \frac{\langle\pd\bar\psi_\gamma(z)\psi_\beta(y) V_m(0)\rangle
\langle \psi_\gamma(z)\bar\psi_\alpha(w)  V_m(0)\rangle}
{\langle V_m(0)\rangle^2}=\\=
-\sum_\gamma\oint_A dz \left(Y(w)^{-1}\Xi(z-w)Y(z)\right)_{\alpha\gamma}\pd_z\left(Y(z)^{-1}\Xi(y-z)Y(y)\right)_{\gamma\beta}=\\=
\sum_\gamma\oint_A dz \pd_z\left(Y(w)^{-1}\Xi(z-w)Y(z)\right)_{\alpha\gamma}\left(Y(z)^{-1}\Xi(y-z)Y(y)\right)_{\gamma\beta}=\\=
\oint_Adz\left(Y(w)^{-1}\left(\Xi(z-w)L(z)+\pd_z\Xi(z-w)\right)\Xi(y-z)Y(y)\right)_{\alpha\beta}.\label{eq:F_final}
}
Comparing \eqref{eq:F_initial} with \eqref{eq:F_final} and using \eqref{eq:LaxPairSystem} we get an identity
\eq{
M(w)\Xi(y-w)-\Xi(y-w)M(y)+2\pi i\pd_\tau\Xi(y-w)=\\=\oint_A dz\left(\Xi(z-w)L(z)+\pd_z\Xi(z-w)\right)\Xi(y-z).
}
We can plug in the explicit expression for $L,M$ and see that the relation will hold iff the following relations are satisfied:
\begin{equation}\label{eq:rel1}
\begin{split}
2\pi i& \partial_\tau x(Q,w-y)=p\partial_Qx(Q,w-y)-\partial_w\partial_Qx(Q,w-y)\\
& =\oint_Adz\left[px(Q,w-z)x(Q,z-y)-\partial_wx(Q,w-z) x(Q,z-y) \right],
\end{split}
\end{equation}
\begin{equation}\label{eq:rel2}
y(2Q,w)x(Q,y-w)-y(2Q,y)x(-Q,y-w)=\oint_Adz x(2Q,z)x(-Q,z-w)x(Q,y-z).
\end{equation} 
To find the r.h.s. we need to compute two integrals:
\eq{
I_1(w,y)=\oint_A dz\ x(Q,w-z)x(Q,z-y),\\
I_2(w,y)=\oint_A dz\ x(q_1,z-w)x(q_2-q_1,z)x(q_2,y-z).
}
$I_1$ and $\pd_w I_1$ contribute to diagonal elements, whereas $I_2$ defines off-diagonal ones.

First we consider the integral over the boundary of the cut torus:
\eq{
I_1^\epsilon=\oint_{\pd\mathbb T}dz\ x(Q,w-z) x(Q+\epsilon,z-y).
}
The function inside the integral is not periodic under $z\rightarrow z+\tau$, but rather acquires a phase $e^{-2\pi i\epsilon}$. If we take the combination of two contour integrals over $A$-cycles shifted by $\tau$ in the opposite directions, they enter with opposite signs times the quasi-periodicity:
\eq{
I_1^\epsilon=(1-e^{-2\pi i\epsilon})\oint_A dz\ x(Q,w-z) x(Q+\epsilon,z-y).
}
On the other hand, this integral can be computed by residues:
\eq{
I_1^\epsilon=2\pi i\left[x(Q+\epsilon,w-y)-x(Q,w-y)\right].
}
Now we take a limit $\epsilon=0$:
\eq{
I_1=2\pi i\lim_{\epsilon\to 0}\frac{x(Q+\epsilon,w-y)-x(Q,w-y)}{1-e^{-2\pi i\epsilon}}=\pd_Q x(Q,w-y).
}
By plugging this result in \eqref{eq:rel1} we see that it is satisfied. The second integral can be computed in the same manner, noting that the integrand now has quasi-periodicity $e^{2\pi i\epsilon}$:
\eq{
I_2^{\epsilon}=\oint_{\pd\mathbb T}dz\ x(q_1,z-w)x(q_2-q_1+\epsilon,z)x(q_2,y-z)=\\
=2\pi i \left(x(q_1,y-w) x(q_2-q_1+\epsilon,y) - x(q_1,-w)x(q_2,y) - x(q_2-q_1+\epsilon,w)x(q_2,y-w)\right)
}
Now we expand this integral up to the first order and find $I_2$:
\eq{
I_2(w,z)=y(2Q,w)x(Q,y-w)-y(2Q,y)x(-Q,y-w)
}
because of which \eqref{eq:rel2} is satisfied.




\section{Other determinantal formulas}\label{App:DetForm}

We report in this appendix some further identities that can be derived from the Fredholm determinant expression \eqref{eq:determinant}, that we recall here for convenience:
\begin{equation}
Z^D(\tau)=-N(a,m,a) q^{a^2+(\sigma+1/2)^2-1/12}e^{2\pi i\rho}\prod_{n=1}^\infty(1-q^n)^{-2\gamma^2}
\det\left(\mathbb{I}+K\right).
\end{equation}
Notice that $\left.K\right|_{\rho\ \mapsto\rho+1/2}=-K$. Combining this observation with \eqref{eq:tau0_tau12} and with periodicity properties of theta functions we find
\eq{
Z_{1/2}^D(\tau)\eta(\tau)^{-1}\theta_3(2\rho|2\tau)=
\frac12N(a,m,a) q^{a^2-1/3}e^{2\pi i\rho}
\left(\det(\mathbb I- K)-\det(\mathbb I+ K)\right),\\
Z_{0}^D(\tau)\eta(\tau)^{-1}\theta_2(2\rho|2\tau)=
\frac12N(a,m,a) q^{a^2-1/3}e^{2\pi i\rho}
\left(\det(\mathbb I- K)+\det(\mathbb I+ K)\right),
}
where we put $\gamma=0$ to simplify the formulas. For arbitrary $\gamma$ everything is the same.

Third, one can substitute $\rho=\frac14$ and $\rho=\frac14+\frac\tau2$ into \eqref{eq:tau0_tau12} in order to cancel each of the two theta functions:
\eq{
Z_{1/2}^D(\tau)\eta(\tau)^{-1}\theta_4(2\tau)=
-i N(a,m,a) q^{a^2-1/3}
\det\left(\mathbb{I}+\left. K\right|_{\rho=\frac14}\right),\\
Z_{0}^D(\tau)\eta(\tau)^{-1}\theta_4(2\tau)=
N(a,m,a) q^{a^2+5/12}
\det\left(\mathbb{I}+\left. K\right|_{\rho=\frac14+\frac\tau2}\right).
}

\end{appendix}
\bibliographystyle{JHEP}
\bibliography{Biblio.bib}

\end{document}